\documentclass{aa}

\usepackage{graphicx}
\usepackage{txfonts}
\usepackage{natbib}

\bibpunct{(}{)}{;}{a}{}{,}

\newcommand{\cygob}{Cyg~OB2}
\newcommand{\tenlac}{10~Lac}
\newcommand{\tausco}{\mbox{$\tau$~Sco}}
\newcommand{\zoph}{\mbox{$\zeta$~Oph}}
\newcommand{\HD}{HD\,}
\newcommand{\fastwind}{{\sc fastwind}}
\newcommand{\cmfgen}{{\sc cmfgen}}
\newcommand{\pikaia}{{\sc pikaia}}

\newcommand{\bra}{Br$\alpha$}
\newcommand{\ha}{H$\alpha$}
\newcommand{\hb}{H$\beta$}
\newcommand{\hg}{H$\gamma$}
\newcommand{\hd}{H$\delta$}
\newcommand{\hi}{\ion{H}{i}}
\newcommand{\hei}{\ion{He}{i}}
\newcommand{\heii}{\ion{He}{ii}}
\newcommand{\heiii}{\ion{He}{iii}}
\newcommand{\kmsec}{\mbox{km\,s$^{-1}$}}
\newcommand{\cmsecsec}{\mbox{cm\,s$^{-2}$}}
\newcommand{\logg}{\mbox{$\log{{g}}$}}
\newcommand{\loggc}{\mbox{$\log{{g}}_{\rm c}$}}
\newcommand{\mdot}{\mbox{$\dot{M}$}}
\newcommand{\msun}{\mbox{$M_{\sun}$}}
\newcommand{\msunyr}{\mbox{$M_{\sun}{\rm yr}^{-1}$}}

\newcommand{\teff}{\mbox{$T_{\rm eff}$}}
\newcommand{\rstar}{\mbox{$R_{\star}$}}
\newcommand{\lstar}{\mbox{$L_{\star}$}}
\newcommand{\lsun}{\mbox{$L_{\sun}$}}
\newcommand{\rsun}{\mbox{$R_{\sun}$}}
\newcommand{\vturb}{\mbox{$v_{\rm turb}$}}
\newcommand{\vinf}{\mbox{$v_{\infty}$}}
\newcommand{\vsini}{\mbox{$v_{\rm r}\sin i$}}
\newcommand{\redchi}{\mbox{$\chi_{\rm red}^2$}}
\newcommand{\redchii}{\mbox{$\chi_{{\rm red}, i}^2$}}
\newcommand{\yhe}{\mbox{$Y_{\rm He}$}}
\newcommand{\Ms}{\mbox{$M_{\rm s}$}}
\newcommand{\Mev}{\mbox{$M_{\rm ev}$}}
\newcommand{\Mv}{\mbox{$M_{V}$}}
\newcommand{\Dmom}{\mbox{$D_{\rm mom}$}}
\newcommand{\magn}{\mbox{$^{\rm m}$}}

\defcitealias{herrero02}{HPN}
\defcitealias{repolust04}{RPH}

\begin{document}

\title{Spectral analysis of early-type stars using a \\ genetic algorithm
based fitting method}

\titlerunning{Spectral analysis using a genetic algorithm based fitting method}

\author{M.\,R. Mokiem\inst{1} \and A. de Koter\inst{1}
\and J. Puls\inst{2}
\and A. Herrero\inst{3,4}
\and F. Najarro\inst{5}
\and M.\,R. Villamariz\inst{3}
}

\authorrunning{M.\,R. Mokiem et al.}

\offprints{M.R. Mokiem, \\ \email{mokiem@science.uva.nl}}
\institute{
  Astronomical Institute Anton Pannekoek, University of Amsterdam,
  Kruislaan 403, 1098~SJ Amsterdam, The Netherlands
  \and
  Universit\"ats-Sternwarte M\"unchen, Scheinerstr. 1, D-81679 M\"unchen,
  Germany
  \and
  Instituto de Astrof\'{\i}sica de Canarias, E-38200, La Laguna,
  Tenerife, Spain
  \and
  Departamento de Astrof\'{\i}sica, Universidad de La Laguna,
  Avda.\ Astrof\'{\i}sico Francisco S\'anchez, s/n, E-38071
  La Laguna, Spain
  \and
  Instituto de Estructura de la Materia, Consejo Superior de
  Investigaciones Cient\'{\i}ficas, CSIC, Serrano 121, E-28006
  Madrid, Spain
}

\date{Received / Accepted}

\abstract{We present the first automated fitting method for the
quantitative spectroscopy of O- and early B-type stars with stellar
winds. The method combines the non-LTE stellar atmosphere code
\fastwind\ from \cite{puls05} with the genetic algorithm based
optimizing routine \pikaia\ from \cite{charbonneau95}, allowing for a
homogeneous analysis of upcoming large samples of early-type stars
\citep[e.g.][]{evans05}. In this first implementation we use continuum
normalized optical hydrogen and helium lines to determine photospheric
and wind parameters. We have assigned weights to these lines
accounting for line blends with species not taken into account,
lacking physics, and/or possible or potential problems in the model
atmosphere code. We find the method to be robust, fast, and
accurate. Using our method we analysed seven O-type stars in the young
cluster \cygob\ and five other Galactic stars with high rotational
velocities and/or low mass loss rates (including \tenlac, \zoph, and
\tausco) that have been studied in detail with a previous version of
\fastwind. The fits are found to have a quality that is comparable or
even better than produced by the classical ``by eye'' method. We
define errorbars on the model parameters based on the maximum
variations of these parameters in the models that cluster around the
global optimum. Using this concept, for the investigated dataset we
are able to recover mass-loss rates down to $\sim$ $6 \times 10^{-8}
\msunyr$ to within an error of a factor of two, ignoring possible
systematic errors due to uncertainties in the continuum
normalization. Comparison of our derived spectroscopic masses with
those derived from stellar evolutionary models are in very good
agreement, i.e.\ based on the limited sample that we have studied we do
not find indications for a mass discrepancy. For three stars we find
significantly higher surface gravities than previously reported. We
identify this to be due to differences in the weighting of Balmer line
wings between our automated method and ``by eye'' fitting and/or an
improved multidimensional optimization of the parameters. The
empirical modified wind momentum relation constructed on the basis of
the stars analysed here agrees to within the error bars with the
theoretical relation predicted by \cite{vink00}, including those cases
for which the winds are weak (i.e.\ less than a few times $10^{-7}$
\msunyr).

  \keywords{methods: data analysis - line: profiles -
            stars: atmospheres - stars: early-type - 
            stars: fundamental parameters - stars: mass loss
  }
}

\maketitle

\section{Introduction} 
Until about a decade ago detailed analysis of the photospheric and
wind properties of O-type stars was limited to about 40 to 50 stars
divided over the Galaxy and the Magellanic Clouds (see e.g.\
\citealt{puls96}; see also \citealt*{repolust04}). The reason that at
that time only such a limited number of objects had been investigated
is related in part to the fact that considerable effort was directed
towards improving the physics of the non-local thermodynamic
equilibrium (non-LTE) model atmospheres used to analyse massive
stars. Notable developments have been the improvements in the atomic
models \citep[e.g.][]{becker92}, shock treatment \citep{pauldrach01},
clumping \citep{hillier91, hillier99}, and the implementation of line
blanketing \citep[e.g.][]{hubeny95, hillier98, pauldrach01}. To study
the effects of these new physics a core sample of ``standard'' O-type
stars has been repeatedly re-analysed. A second reason, that is at
least as important, is the complex, and time and CPU intensive nature
of these quantitative spectroscopic analyses. Typically, at least a
six dimensional parameter space has to be probed, i.e.\ effective
temperature, surface gravity, helium to hydrogen ratio, atmospheric
microturbulent velocity, mass-loss rate, and a measure of the
acceleration of the transonic outflow.  Rotational velocities and
terminal outflow velocities can be determined to considerable accuracy
by means of external methods such as rotational (de-) convolution
methods \citep[e.g.][]{howarth97} and SEI-fitting of P-Cygni lines
\citep[e.g.][]{groenewegen89}, respectively. To get a good spectral
fit it typically requires tens, sometimes hundreds of models per
individual star.

In the last few years the field of massive stars has seen the
fortunate development that the number of O-type stars that have been
studied spectroscopically has been doubled \citep[e.g.][]{crowther02,
herrero02, bianchi02, bouret03, hillier03, garcia04, martins04,
massey04, evans04}. The available data set of massive O- and early
B-type stars has recently {\em again} been doubled, mainly through the
advent of multi-object spectroscopy. Here we explicitly mention the
{\em VLT-FLAMES Survey of Massive Stars} \citep{evans05} comprising
over 100 hours of VLT time.  In this survey multi-object spectroscopy
using the {\em Fibre Large Array Multi-Element Spectrograph} (FLAMES)
has been used to secure over 550 spectra (of which in excess of 50 are
spectral type O) in a total of seven clusters distributed over the
Galaxy and the Magellanic Clouds.

This brings within reach different types of studies that so far could
only be attempted with a troublingly small sample of stars. These
studies include establishing the mass loss behaviour of Galactic stars
across the upper Hertzsprung-Russell diagram, from the weak winds of
the late O-type dwarfs (of order $10^{-8}$~\msunyr) to the very strong
winds of early O-type supergiants (of order $10^{-5}$~\msunyr);
determination of the mass-loss versus metallicity dependence in the
abundance range spanned by Small Magellanic Cloud to Galactic stars;
placing constraints on the theory of massive star evolution by
comparing spectroscopic mass determinations and abundance patterns
with those predicted by stellar evolution computations, and the study
of (projected) spatial gradients in the mass function of O- and B-type
stars in young clusters, as well as such spatial gradients in the
initial atmospheric composition of these stars.

To best perform studies such as listed above not only requires a large
set of young massive stars, it also calls for a robust, homogeneous
and objective means to analyse such datasets using models that include
state-of-the-art physics. This essentially requires an automated
fitting method. Such an automated method should not only be fast, it
must also be sufficiently flexible to be able to treat early-type
stars with widely different properties (e.g. mass-loss rates that
differ by a factor of $10^{3}$). Moreover, it should apply a well
defined fitting criterium, like a $\chi^{2}$ criterium, allowing it to
work in an automated and reproducible way.

To cope with the dataset provided by the {\em VLT-FLAMES Survey} and
to improve the objectivity of the analysis, we have investigated the
possibility of automated fitting. Here we present a robust, fast, and
accurate method to perform automated fitting of the continuum
normalized spectra of O- and early B-type stars with stellar winds
using the fast performance stellar atmosphere code \fastwind\
\citep{puls05} combined with a genetic algorithm based fitting method.
This first implementation of an automated method should therefore be
seen as an improvement over the standard ``by eye'' method, and not as
a replacement of this method. The improvement lies in the fact that
with the automated method large data sets (tens or more stars),
spanning a wide parameter space, can be analysed in a repeatable and
homogeneous way. It does not replace the ``by eye'' method as our
automated fitting method still requires a by eye continuum
normalization as well as a human controlled line selection. This
latter should address the identification and exclusion of lines that
are not modeled (i.e.\ blends), as well as introduce information on
lacking physics and/or possible or potential problems in the model
atmosphere code.  Future implementations of an automated fitting
method may use the absolute spectrum, preferably over a broad
wavelength range. This would eliminate the continuum rectification
problem, however, it will require a modeling of the interstellar
extinction. In this way one can work towards a true replacing of the
``by eye'' method by an automated approach.

In Sect.~\ref{sec:auto_fit} we describe the genetic algorithm method
and implementation, and we provide a short r\'{e}sum\'{e} of the
applied unified, non-LTE, line-blanketed atmosphere code \fastwind\ --
which is the only code to date for which the method described here is
actually achievable (in the context of analysing large data sets). To
test the method we analyse a set of 12 early type spectra in
Sect.~\ref{sec:spec-analysis}. We start with a re-analysis of a set of
seven stars in the open cluster \object{\cygob} that have been studied
by \cite{herrero02}. The advantage of focusing on this cluster is that
it has been analysed with a previous version of \fastwind, allowing
for as meaningful a comparison as is possible, while still satisfying
our preference to present a state-of-the-art analysis. The analysis of
\cygob\ has the added advantage that all stars studied are
approximately equidistant. To test the performance of our method
outside the parameter range offered by the \cygob\ sample we have
included an additional five well-studied stars with either low density
winds and/or very high rotational velocities. In
Sect.~\ref{sec:errors} we describe our error analysis method for the
multidimensional spectral fits obtained with the automated method. A
systematic comparison of the obtained parameters with previously
determined values is given in Sect.~\ref{sec:comp}. Implications of
the newly obtained parameters on the properties of massive stars are
discussed in Sect.~\ref{sec:implic}. In the last section we give our
conclusions.

\section{Automated fitting using a genetic algorithm}
\label{sec:auto_fit}

\subsection{Spectral line fitting as an optimization problem}
\label{sec:opt_prob}
Spectral line fitting of early-type stars is an optimization problem
in the sense that one tries to maximize the correspondence between a
given observed spectrum and a synthetic spectrum produced by a stellar
atmosphere model. Formally speaking one searches for the global
optimum, i.e.\ best fit, in the parameter space spanned by the free
parameters of the stellar atmosphere model by minimizing the
differences between the observed and synthesized line profiles.

Until now the preferred method to achieve this minimization has been
the so called fitting ``by eye'' method. In this method the best fit
to the observed spectrum of a certain object is determined in an
iterative manner. Starting with a first guess for the model parameters
a spectrum is synthesized. The quality of the fit to the observed
spectrum is determined, as is obvious from the methods name, by an
inspection by eye. Based on what the person performing the fit sees,
for instance, whether the width of the line profiles are reproduced
correctly, combined with his/her experience and knowledge of the
model and the object, the model parameters are modified and a new
spectrum is synthesized. This procedure is repeated until the quality
of the fit determined by eye cannot be increased anymore by modifying
the model parameters.

It can be questioned whether a fit constructed with the fitting ``by
eye'' method corresponds to the best fit possible, i.e.\ the global
optimum. Reasons for this are, {\em i)} the restricted size of
parameter space that can be investigated, both in terms of number of
free parameters as well as absolute size of the parameter domain that
can investigated with high accuracy, {\em ii)} the limited number of
free parameters that are changed simultaneously, and {\em iii)} biases
introduced by judging the quality of a line fit by eye. The importance
of the first point lies in the fact that in order to assure that the
global optimum is found, a parameter space that is as large as
possible should be explored with the same accuracy for all parameters
in the complete parameter space. If this is not the case the solution
found will likely correspond to a local optimum.

The argument above becomes stronger in view of the second point.
Spectral fitting is a multidimensional problem in which the line
profile shapes depend on all free parameters simultaneously, though to
a different extent.  Consequently, the global optimum can only be
found if all parameters are allowed to vary at the same time. The use
of fit diagrams \cite[e.g.][]{kudritzki78, herrero92} does not resolve
this issue. These diagrams usually only take variations in \teff\ and
\logg\ into account, neglecting the effects of other parameters, like
microturbulence \cite[e.g.][]{smith98, villamariz00} and mass loss
(e.g. [Fig.\ 5 of] \citealt{mokiem04}), on the line profiles.

The last point implies that, strictly speaking, fitting ``by eye''
cannot work in a reproducible way. There is no uniform well defined
method to judge how well a synthetic line profile fits the data by
eye. More importantly, it implies that there is no guarantee that the
synthetic line profiles selected by the eye, correspond to the
profiles which match the data the best. This predominantly increases
the uncertainty in the derivation of those parameters that very
sensitively react to the line profile shape, like for instance the
surface gravity.

The new fitting method presented here does not suffer from the
drawbacks discussed above. It is an automated method capable of global
optimization in a multi-dimensional parameter space of arbitrary size
(Sect.~\ref{sec:fit-param}). As it is automated, it does not require
any human intervention in finding the best fit, avoiding potential
biases introduced by ``by eye'' interpretations of line profiles. The
method described here consists of two main components. The first
component is the non-LTE stellar atmosphere code
\fastwind. Section~\ref{sec:FastWind} gives an overview of the
capabilities of the code and the assumptions involved. The second
component is the genetic algorithm (GA) based optimizing routine
\pikaia\ from \cite{charbonneau95}, which is responsible for
optimizing the parameters of the \fastwind\ models. For the technical
details of this routine and more information on GAs we refer to the
cited paper and references therein. Here we will suffice with a short
description of GAs and a description of the GA implementation with
respect to optimization of spectral fits.

\subsection{The genetic algorithm implementation}
Genetic algorithms represent a class of heuristic optimization
techniques, which are inspired by the notion of evolution by means of
natural selection \citep{darwin1859}. They provide a method of solving
optimization problems by incorporating this biological notion in a
numerical fashion. This is achieved by evolving the global solution
over subsequent generations starting from a set of randomly guessed
initial solutions, so called individuals. Selection pressure is
imposed in between generations based on the quality of the solutions,
their so called fitness. A higher fitness implies a higher probability
the solution will be selected for reproduction. Consequently, only a
selected set of individuals will pass on their ``genetic material'' to
subsequent new generations.

To create the new generations discussed above GAs require a
reproduction mechanism. In its most basic form this mechanism consists
of two genetic operators. These are the crossover operator, simulating
sexual reproduction, and the mutation operator, simulating copying
errors and random effects affecting a gene in isolation. An important
benefit of these two operators is the fact that they also introduce
new genetic material into the population. This allows the GA to
explore new regions of parameters space, which is important in view of
the existence of local extremes. When the optimization runs into a
local optimum, these two operators, where usually mutation has the
strongest effect, allow for the construction of individuals outside of
this optimum, thereby allowing it to find a path out of the local
optimum. This capability to escape local extremes, consequently,
classifies GAs as global optimizers and is one of the reasons they
have been applied to many problems in and outside astrophysics
\citep[e.g.][]{metcalfe00, gibson98}.

Using an example we can further illustrate the GA optimization
technique. Lets assume that the optimization problem is the
minimization of some function $f$. This function has $n$ variables,
serving as the genetic building blocks, spanning a $n$ dimensional
parameter space. The first step in solving this problem is to create
an initial population of individuals, which are sets of $n$
parameters, randomly distributed in parameter space. For each of these
individuals the quality of their solution is determined by simply
calculating $f$ for the specific parameter values. Now selection
pressure is imposed and the fittest individuals, i.e.\ those that
correspond to the lowest values of $f$, are selected to construct a
new generation. As the selected individuals represent the fittest
individuals from the population, every new generation will consist of
fitter individuals, leading to a minimization of $f$, thereby, solving
the optimization problem.

With the previous example in mind we can explain our implementation of
the GA for solving the optimization problem of spectral line fitting,
with the following scheme. We start out with a first generation of a
population of \fastwind\ models randomly distributed in the free
parameter space (see Sect.~\ref{sec:fit-param}). For each of these
models it is determined how well an observed spectrum is fitted by
calculating the reduced chi squared, \redchii, for each of the fitted
lines $i$. The fitness $F$, of a model is then defined as the inverted
sum of the \redchii's, i.e.
\begin{equation}
  \label{eq:fitns}
  F \equiv \left(\sum_i^N \chi_{{\rm red}, i}^2\right)^{-1}~,
\end{equation}
where $N$ corresponds to the number of lines evaluated. The fittest
models are selected and a new generation of models is constructed
based on their parameters. From this generation the fitnesses of the
models are determined and again from the fittest individuals a new
generation is constructed. This is repeated until $F$ is maximized,
i.e.\ a good fit is obtained.

In terms of quantifying the fit quality Eq.~(\ref{eq:fitns}) does not
represent a unique choice. Other expressions for the fitness
criterium, for instance, the sum of the inverted \redchii's of the
individual lines, or the inverted \redchi\ of all the spectral points
evaluated, also produce the required functionality of an increased
fitness with an increased fit quality. We have chosen this particular
form based on two of its properties. Firstly, the evaluation of the
fit quality of the lines enter into the expression individually,
ensuring that, regardless of the number of points in a certain line,
all lines are weighted equally. This allows as well for weighting
factors for individual lines, which express the quality with which the
stellar atmosphere synthesizes these lines (cf.\
Sect.~\ref{sec:line-scheme}). Secondly, using the inverted sum of the
\redchii's instead of the sum of the inverted \redchii's avoids having
a single line, which is fitted particularly well, to dominate the
solution. Instead the former form demands a good fit of all lines
simultaneously.

\subsection{Parallelization of the genetic algorithm}
The ability of global optimization of GAs comes at a price. Finding
the global minimum requires the calculation of many generations. In
Sect.~\ref{sec:formal-tests} we will show that for the spectra studied
in this paper, the evaluation of more than a hundred generations is
needed to assure that the global optimum is found. For a typical
population size of $\sim$70 individuals, this comes down to the
calculation of $\sim$7000 \fastwind\ models. With a modern 3~GHz
processor a single \fastwind\ model (aiming at the analysis of
hydrogen and helium lines) can be calculated within five to ten
minutes. Consequently, automated fitting on a sequential computer
would be unworkable.

To overcome this problem, parallelization of the \pikaia\ routine is
necessary. This parallelization is inspired by the work of
\cite{metcalfe03}. Consequently, our parallel version is very similar
to the version of these authors. The main difference between the two
versions, is an extra parallelization of the so called elitism option
in the reproduction schemes (see \citeauthor{metcalfe03}). This was
treated in a sequential manner in the \citeauthor{metcalfe03}
implementation and has now been parallelized as well.

Due to the strong inherent parallelism of GAs, the parallel version of
our automated fitting method scales very well with the number of
processors used. Test calculations showed that for configurations in
which the population size is an integer multiple of the number of
processors the sequential overhead is negligible. Consequently, the
runtime scales directly with the inverse of the number of
processors. Thus enabling the automated fitting of spectra.

\subsection{The non-LTE model atmosphere code \fastwind}
\label{sec:FastWind}
For modeling the optical spectra of our stars we use the latest
version of the non-LTE, line-blanketed atmosphere code \fastwind\ for
early-type stars with winds. For a detailed description we refer to
\cite{puls05}. Here we give a short overview of the assumptions made
in this method. The code has been developed with the emphasis on a
{\em fast performance} (hence its name), which makes it currently the
best suited (and realistically only) model for use in this kind of
automated fitting methods.

\fastwind\ adopts the concept of ``unified model atmospheres'',
i.e.\ including both a pseudo-hydrostatic photosphere and a transonic
stellar wind, assuring a smooth transition between the two. The
photospheric density structure follows from a self-consistent solution
of the equation of hydrostatic equilibrium and accounts for the actual
temperature stratification and radiation pressure. The temperature
calculation utilizes a flux-correction method in the lower atmosphere
and the thermal balance of electrons in the outer atmosphere (with a
lower cut-off at $T_{\rm min} = 0.5 \teff$). In the photosphere the
velocity structure, $v(r)$, corresponds to quasi-hydrostatic
equilibrium; outside of this regime, in the region of the sonic
velocity and in the super-sonic wind regime it is prescribed by a
standard $\beta$-type velocity law, i.e.
\begin{equation}
   v(r) = \vinf \left( 1 - \frac{r_{\circ}}{r} \right)^{\beta}~,
\end{equation}
where \vinf\ is the terminal velocity of the wind. The parameter
$r_{\circ}$ is used to assure a smooth connection, and $\beta$ is a
measure of the flow acceleration.

The code distinguishes between {\em explicit} elements (in our case
hydrogen and helium) and {\em background} elements (most importantly:
C, N, O, Ne, Mg, Si, S, Ar, Fe, Ni). The explicit elements are used as
diagnostic tools and are treated with high precision, i.e.\ by detailed
atomic models and by means of {\em co-moving-frame} transport for the
line transitions. The \hi\ and \heii\ model atoms consist of 20 levels
each; the \hei\ model includes levels up to and including $n = 10$,
where levels with $n \ge 8$ have been packed.  The background ions are
included to allow for the effects of line-blocking (treated in an
approximate way by using suitable means for the corresponding line
opacities) and line-blanketing. Occupation numbers and opacities of
both the explicit and the most abundant background ions are
constrained by assuming statistical equilibrium. The only difference
between the treatment of these types of ions is that for the
background ions the Sobolev approximation is used in describing the
line transfer (accounting for the actual illumination radiation
field).

Abundances of the background elements are taken from the solar values
provided by \citet[][and references therein]{grevesse98}. The He/H
ratio is not fixed and can be scaled independently from the background
element abundances.

A comparison between the optical H and He lines as synthesized by
\fastwind\ and those predicted by the independent comparison code
\cmfgen\ \citep{hillier98} show excellent agreement, save for the
\hei\ singlet lines in the temperature range between 36\,000 and
41\,000~K for dwarfs and between 31\,000 and 35\,000~K for
supergiants, where \cmfgen\ predicts weaker lines. We give account of
this discrepancy, and therefore of an increased uncertainty in the
reproduction of these lines, by introducing weighting factors, which
for the \hei\ singlets of stars in these ranges are lower (cf.\
Sect.~\ref{sec:line-scheme}).

\begin{table*}[t]
  \caption{Input parameters of the formal test models (``In'' column)
  and parameters obtained with the automated fitting method by fitting
  synthetic data created from these models (``Out'' column). Results
  were obtained by evolving a population of 72 \fastwind\ models over
  200 generations.}
  \label{tab:form-tests}
  \begin{center}
  \begin{tabular}{lrrrrrrrrr}
  \hline\\[-9pt] \hline \\[-7pt]
                            & Set A& Search       &      & Set B & Search       &      & Set C& Search              \\[2pt]
                            & In   & range        & Out  & In    & range        & Out  & In   & range        & Out  \\[1pt]
  \hline \\[-9pt]
  Spectral type             & O3~I &              &      & O5.5~I&              &      & B0~V &                     \\[3.5pt]
  \teff\ [kK]               & 45.0 & [42, 47]     & 45.0 & 37.5  & [35, 40]     & 37.6 & 30.0 & [28, 34]     & 29.9 \\[3.5pt]
  \logg\ [\cmsecsec]        & 3.80 & [3.5, 4.0]   & 3.84 & 3.60  & [3.3, 3.9]   & 3.57 & 4.00 & [3.7, 4.3]   & 3.95 \\[3.5pt]
  \rstar\ [\rsun]           & 17.0 &              &      & 20.0  &              &      & 8.0  &              &      \\[3.5pt]
  $\log \lstar$ [\lsun]     & 6.03 &              & -    & 5.85  &              & -    & 4.67 &              & -    \\[3.5pt]
  \vturb\ [\kmsec]          & 5.0  & [0, 20]      & 5.9  & 10.0  & [0, 20]      & 9.7  & 15.0 & [0, 20]      & 14.8 \\[3.5pt]
  \yhe                      & 0.15 & [0.05, 0.30] & 0.15 & 0.10  & [0.05, 0.30] & 0.10 & 0.10 & [0.05, 0.30] & 0.10 \\[3.5pt]
  \mdot\ [$10^{-6}$\msunyr] & 10.0 & [1.0, 20.0]  & 9.3  & 5.0   & [1.0, 10.0]  & 5.3  & 0.01 & [0.001, 0.2] & 0.008\\[3.5pt]
  $\beta$                   & 1.20 & [0.5, 1.5]   & 1.18 & 1.00  & [0.5, 1.5]   & 0.99 & 0.80 & [0.5, 1.5]   & 0.93 \\[3.5pt]
  \vinf\ [\kmsec]           & 2500 &              & -    & 2200  &              & -    & 2000 &              & -    \\[3.5pt]
  \vsini\ [\kmsec]          & 150  &              & -    & 120   &              & -    & 90   &              & -    \\[1pt]
  \hline
  \end{tabular}
  \end{center}
\end{table*}

\subsection{Fit parameters}
\label{sec:fit-param}
The main parameters which will be determined from a spectral fit using
\fastwind\ are the effective temperature \teff, the surface gravity
$g$, the microturbulent velocity \vturb, the helium over hydrogen
number density \yhe, the mass loss rate \mdot\ and the exponent of the
beta-type velocity law $\beta$. These parameters span the free
parameter space of our fitting method. The stellar radius, \rstar, is
not a free parameter as its value is constrained by the absolute
visual magnitude \Mv. To calculate \rstar\ we adopt the procedure
outlined in \cite{kudritzki80}, i.e.
\begin{equation}
  5\log R/R_{\sun} = 29.57 - (\Mv - V)~,
\end{equation}
where $V$ is the visual flux of the theoretical model given by
\begin{equation}
  -2.5 \log \int_0^\infty F_\lambda S_\lambda d\lambda~.
\end{equation}
In the above equation $S_\lambda$ is the $V$-filter function of
\cite{matthews63} and $F_\lambda$ is the theoretical stellar
flux. Note that as \rstar\ is an input parameter, $F_\lambda$ is not
known before the \fastwind\ model is calculated. Therefore, during the
automated fitting we approximate $F_\lambda$ by a black body radiating
at $T=0.9\teff$ \citep[cf.][]{markova04}. After the fit is completed
we use the theoretical flux from the best fit model to calculate the
non approximated stellar radius. Based on this radius we rescale the
mass loss rate using the invariant wind-strength parameter $Q$
\citep{puls96, dekoter97}
\begin{equation}
  Q = \frac{\mdot}{\left(\vinf \rstar\right)^\frac{3}{2}}~.
\end{equation}
The largest difference between the approximated and final stellar
radius for the objects studied here, is $\sim$2 percent. The
corresponding rescaling in \mdot\ is approximately three percent.

The projected rotation velocity, \vsini, and terminal velocity of the
wind are not treated as free parameters. The value of \vsini\ is
determined from the broadening of weak metal lines and the width of
the \hei\ lines. For \vinf\ we adopt values obtained from the study of
ultraviolet (UV) resonance lines, or, if not available, values from
calibrations are used.

Our fitting method only requires the size of the free parameter domain
to be specified. For the objects studied in this paper we keep the
boundaries between which the parameters are allowed to vary, fixed for
\vturb, \yhe\ and $\beta$. The adopted ranges, respectively, are [0,
20] \kmsec, [0.05, 0.30] and [0.5, 1.5]. The boundaries for \teff\ are
set based on the spectral type and luminosity class of the studied
object. Usually the size of this range is set to approximately
5000~K. The \logg\ range is delimited so that the implied stellar mass
lies between reasonable boundaries. For instance for the B1\,I star
\cygob~\#2 the adopted \teff\ range together with its absolute visual
magnitude imply a possible range in \rstar\ of [11.5:12.0]\rsun. For
the automated fit we set the minimum and maximum \logg\ to 3.1 and
3.8, respectively, which sets the corresponding mass range that will
be investigated to [5.0:25.2]~\msun. For the mass loss rate we adopt a
conservative range of at least one order of magnitude. As example for
the analysis of \cygob~\#2 we adopted lower and upper boundaries of
$4\times 10^{-8}$ and $2\times 10^{-6}\,\msunyr$, respectively.

\begin{figure*}[t]
\centering
  \resizebox{14cm}{!}{\includegraphics{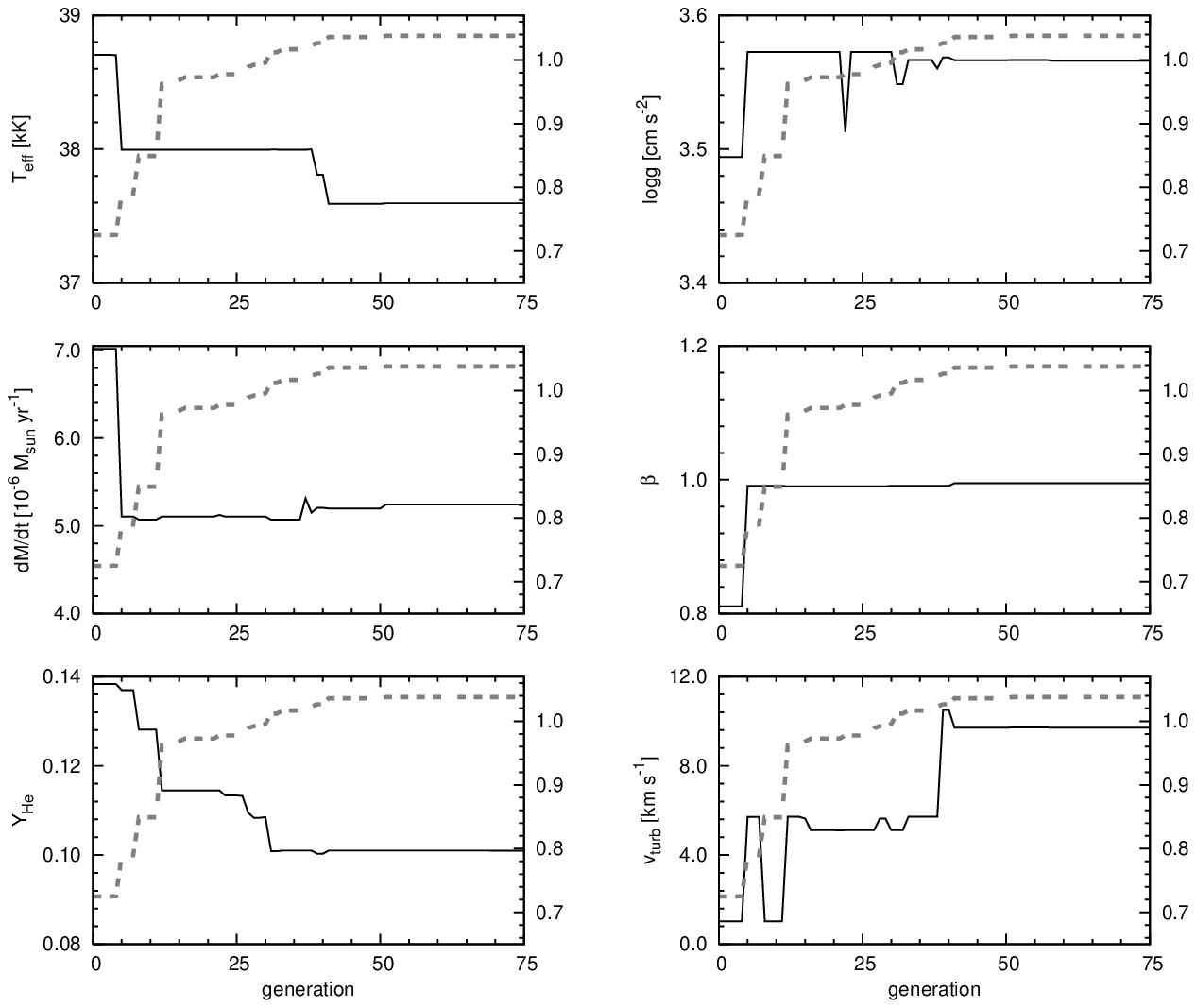}}
  \caption{Evolution of the best fitting model parameters for formal
  test B. From the 200 generation run only the first 75 generations
  are shown. For this specific data set the location of the global
  optimum is found within 50 generations. This is indicated by the
  highest fitness found during the run, which is shown as a grey
  dashed line and is scaled to the right vertical axis. The fitness is
  normalized with respect to the fitness of the model used to create
  the synthetic data (the data being this model plus noise).}
  \label{fig:param_form}
\end{figure*}

\subsection{Formal tests of convergence}
\label{sec:formal-tests}
Before we apply our automated fitting method to real spectra, we first
test whether the method is capable of global optimization. For this we
perform convergence tests using synthetic data. The main goal of these
tests is to determine how well and how fast the input parameters, used
to create the synthetic data, can be recovered with the method.  The
speed with which the input parameters are recovered, i.e.\ the number
of generations needed to find the global optimum, can then be used to
determine how many generations are needed to obtain the best fit for a
real spectrum. In other words, when the fit has converged to the
global optimum.

Three synthetic datasets, denoted by A, B and C, were created with the
following procedure. First, line profiles of Balmer hydrogen lines and
helium lines in the optical blue and \ha\ in the red calculated by
\fastwind\ were convolved with a rotational broadening profile. Table
\ref{tab:form-tests} lists the parameters of the three sets of models
as well as the projected rotational velocity used. A second
convolution with a Gaussian instrumental profile was applied to obtain
a spectral resolution of 0.8~\AA\ and 1.3~\AA\ for, respectively, the
\ha\ line and all other lines. These values correspond to the minimum
resolution of the spectra fitted in Sect.~\ref{sec:spec-analysis}.
Finally, Gaussian distributed noise, corresponding to a signal to
noise value of 100, was added to the profiles.  Dataset A represents
an O3~I star with a very dense stellar wind
($\mdot=10^{-5}\,\msunyr$), while set B is that of an O5.5~I with a
more typical O-star mass loss.  The last set C is characteristic for a
B0~V star with a very tenuous wind of only $10^{-8}\,\msunyr$.

From the synthetic datasets we fitted nine lines, three hydrogen,
three neutral helium and three singly ionized helium lines,
corresponding to the minimum set of lines fitted for a single object
in Sect.~\ref{sec:spec-analysis}. The fits were obtained by evolving a
population of 72 \fastwind\ models over a course of 200
generations. In this test and throughout the remainder of the paper we
use \pikaia\ with a dynamically adjustable mutation rate, with the
minimum and maximum mutation rate set to the default values (see
\citealt{charbonneau95b}). Selection pressure, i.e.\ the weighting of
the probability an individual will be selected for reproduction based
on its fitness, was also set to the default value.

Table~\ref{tab:form-tests} lists the parameter ranges in
which the method was allowed to search, i.e.\ the minimum and maximum
values allowed for the parameters of the \fastwind\ models. As \vinf\
and \vsini\ are not free parameters these were set equal to the input
values.

In all the three test cases the automated method was able to recover
the global optimum. Table~\ref{tab:form-tests} lists the parameters of
the best fit models obtained by the method in the ``Out''
columns. Compared to the parameters used to create the synthetic data,
there is very good agreement. Moderate differences (of a 15-20\%
level) are found for \vturb\ recovered from dataset A and for the wind
parameters $\beta$ and \mdot\ recovered from dataset C. This was to be
expected. In the case of the wind parameters the precision with which
information about these parameters can be recovered from the line
profiles decreases with decreasing wind density
\citep[e.g.][]{puls96}. Still, the precision with which the wind
parameters are recovered for the weak wind data set C, is remarkable.

A similar reasoning applies for the microturbulent velocity recovered
from data set A. For low values of the microturbulence, i.e.\ $\vturb
< v_{\rm th}$, thermal broadening will dominate over broadening due to
microturbulence. This decreases the precision with which this
parameter can be recovered from the line profiles. Realizing that in
case of this dataset for helium $v_{\rm th}~\approx~14~\kmsec$, again,
the precision with which \vturb\ is recovered, is impressive.

To illustrate how quickly and how well the input parameters are
recovered Fig.~\ref{fig:param_form} shows the evolution of the fit
parameters during the fit of synthetic dataset B. Also shown, as a
grey dashed line, is the fitness of the best fitting model found,
during the run. This fitness is normalized with respect to the fitness
of the model used to create the synthetic data (the data being the
combination of this model and noise). Note that the final maximum
normalized fitness found by the method exceeds 1.0, which is due to
the added noise allowing a further fine tuning of the parameters by
the GA based optimization. As can be seen in this figure the method
modifies multiple parameters simultaneously to produce a better
fit. This allows for an efficient exploration of parameter space and,
more importantly, it allows for the method to actually find the global
optimum. 

In the case of dataset B finding the global optimum required only
a few tens of generations ($\sim$30). For the other two datasets all
save one parameter were well established within this number of
generations. To establish the very low value of \vturb\ in dataset A
and the very low \mdot\ in dataset C required $\sim$100
generations. We will adopt 150 generations to fit the spectra in
Sect.~\ref{sec:spec-analysis}. One reason, obviously, is that to
safeguard that the global optimum is found. A second reason, however,
is that it assures that the errors on the model parameters that we
determine are meaningful (i.e.\ it assures that the error on the error
is modest).

We consider doing such a formal test as performed above as part of the
analysis of a set of observed spectra, as the exact number of
generations required is, in principle, a function of e.g.\ the
signal-to-noise ratio and the spectral resolution. Also, special
circumstances may play a role, such as potential nebular contamination
(in which case the impact of removing the line cores from the fit
procedure needs to be assessed).

\begin{table*}
  \caption{Basic parameters of the early type stars studied
  here. Spectral types are taken from \cite{massey91},
  \cite{walborn72, walborn73} and \cite{conti71}. Blue and red
  resolution, respectively, correspond to the region between
  $\sim$4000 and $\sim$5000~\AA\ and the region around \ha.}
  \label{tab:data}
  \begin{center}
  \begin{tabular}{llcccrc}
    \hline\\[-9pt] \hline \\[-7pt]
  Star & Spectral & \Mv & Blue & Red & \multicolumn{1}{c}{\vsini} & \vinf\\[2pt]
       & Type & & resolution [\AA] & resolution [\AA] & [\kmsec] & [\kmsec] \\[1pt]
  \hline\\[-9pt]
  \object{\cygob~\#7}  & O3~If$^*$   & $-5.91$ & 0.6 & 0.8 & 105 & 3080\\[3.5pt]
  \object{\cygob~\#11} & O5~If$^+$   & $-6.51$ & 1.3 & 0.8 & 120 & 2300\\[3.5pt]
  \object{\cygob~\#8C} & O5~If       & $-5.61$ & 1.3 & 0.8 & 145 & 2650\\[3.5pt]
  \object{\cygob~\#8A} & O5.5~I(f)   & $-6.91$ & 0.6 & 0.8 & 130 & 2650\\[3.5pt]
  \object{\cygob~\#4}  & O7~III((f)) & $-5.44$ & 1.3 & 0.8 & 125 & 2550\\[3.5pt]
  \object{\cygob~\#10} & O9.5~I      & $-6.86$ & 0.6 & 0.8 & 95  & 1650\\[3.5pt]
  \object{\cygob~\#2}  & B1~I        & $-4.64$ & 0.6 & 0.8 & 50  & 1250\\[3.5pt]
  \object{\HD15629}    & O5~V((f))   & $-5.50$ & 0.6 & 0.8 & 90  & 3200\\[3.5pt]
  \object{\HD217086}   & O7~Vn       & $-4.50$ & 0.6 & 0.8 & 350 & 2550\\[3.5pt]
  \object{\tenlac}     & O9~V        & $-4.40$ & 0.6 & 0.6 & 35  & 1140\\[3.5pt]
  \object{\zoph}       & O9~V        & $-4.35$ & 0.6 & 0.8 & 400 & 1550\\[3.5pt]
  \object{\tausco}     & B0.2~V      & $-3.10$ & 0.2 & 0.2 & 5   & 2000\\[1pt]
  \hline
  \end{tabular}
  \end{center}
\end{table*}

\section{Spectral analysis of early-type stars}
\label{sec:spec-analysis}
In this section we apply our fitting method to seven stars in the open
cluster \cygob, previously analysed by \cite{herrero02} and five
``standard'' early-type stars, \tenlac, \tausco, \zoph, \HD15629\ and
\HD217086, previously analysed by various authors.

\subsection{Description of the data}

Table~\ref{tab:data} lists the basic properties of the data used for
the analysis. All spectra studied have a S/N of at least 100. The
spectral resolution of the data in the blue (regions between
$\sim$4000 and $\sim$5000~\AA) and the red (region around \ha) is
given in Tab.~\ref{tab:data}.

The optical spectra of the stars in \cygob\ were obtained by
\cite{herrero99} and \cite{herrero00}. Absolute visual magnitudes of
the \cygob\ objects were adopted from \cite{massey91}, and correspond
to a distance modulus of 11.2\magn. Note that for object \#8A Tab.~7
in \citeauthor{massey91} contains an incorrect $V_0$ value of
4.08\magn. This should have been 4.26\magn\ conform the absorption
given in this table and the visual magnitude in their Tab.~2. For
\vsini\ values determined by \cite{herrero02} are used, with the
exception of objects \#8A and \#10. For these we found that the \hei\
and metal lines are somewhat better reproduced if we adopt \vsini\
that are higher by $\sim$35\% and $\sim$10\%, respectively.  Terminal
flow velocities of the wind have been obtained from UV spectra
obtained with {\em Hubble Space Telescope}
\citep[cf.][]{herrero01}. Data of \HD15629, \HD217086 and \zoph\ are
from \cite{herrero92} and \cite{herrero93}. For \Mv, \vinf\ and
\vsini\ values given by \cite{repolust04} are adopted. The distances
to these objects are based on spectroscopic parallaxes, except for
\zoph\ which has a reliable {\em Hipparcos} distance
\citep{schroder04}.

The spectrum of \tenlac\ was obtained by \cite{herrero02}. The
absolute visual magnitude of this star is from \cite{herrero92}. For
\vinf\ we adopted the minimum value which is approximately equal to
the escape velocity at the stellar surface of this object. For the
projected rotational velocity we adopt 35 \kmsec.
The blue spectrum of \tausco\ is from \cite{kilian92}. The red region
around \ha\ was observed by \cite{zaal99}. For \tausco\ we also adopt
the {\em Hipparcos} distance. This distance results in an absolute
visual magnitude which is rather large for the spectral type of this
object, but is in between the \Mv\ adopted by \cite{kilian92} and
\cite{humphreys78}. For the projected rotational velocity a value of 5
\kmsec\ was adopted.

\subsection{Lines selected for fitting and weighting scheme}
\label{sec:line-scheme}
For the analysis \fastwind\ will fit the hydrogen and helium spectrum
of the investigated objects. Depending on the wavelength range of the
available data, these lines comprise for hydrogen the Balmer
lines \ha, \hb, \hg\ and \hd; for \hei\ the singlet lines at 4387 and
4922\,\AA, the \hei\ triplet lines at 4026, which is blended with
\heii, 4471 and 4713\,\AA; and finally for \heii\ the lines at 4200,
4541 and 4686\,\AA.


For an efficient and reliable use of the automated method we have to
incorporate into it the expertise that we have developed in the
analysis of OB stars. The method has to take into account that some
lines may be blended or that they cannot be completely reproduced by
the model atmosphere code for whatever reason For example, the
so-called ``generalized dilution effect'' \cite{voels89}, present in
the \hei\,$\lambda$4471 line in late type supergiants, that is still
lacking an explanation.

\begin{table}
  \caption{Line weighting scheme adopted for different spectral
   types and luminosity classes for the objects fitted in this
   paper. Late, mid and early spectral type correspond to,
   respectively, [O2-O5.5], [O6-O7.5] and [O8-B1]. The weights are
   implemented in the fitness definition according to
   Eq.~(\ref{eq:fitns_weights}) and have values of 1.0, 0.5 and 0.25
   in case of h, m and l, respectively.}
  \label{tab:line-weights}
  \begin{center}
  \begin{tabular}{lccccccc}
  \hline\\[-9pt] \hline \\[-7pt]
   & \multicolumn{3}{c}{Dwarfs}
   & & \multicolumn{3}{c}{Super Giants} \\[2pt]
   & Late & Mid & Early & & Late & Mid & Early\\[1pt]
 \hline \\[-9pt]
  H Balmer        & h & h & h & & h & h & h \\[3.5pt]
  \hei\  singlets & h & l & l & & h & l & l \\[3.5pt]
  \hei\  4026     & h & h & h & & h & h & h \\[3.5pt]
  \hei\  4471     & h & h & h & & l & m & h \\[3.5pt]
  \hei\  4713     & h & h & h & & h & h & h \\[3.5pt]
  \heii\ 4686     & h & m & m & & m & m & m \\[3.5pt]
  \heii\ 4541     & h & h & h & & h & h & h \\[3.5pt]
  \heii\ 4200     & m & m & m & & m & m & m \\[1pt]
  \hline
  \end{tabular}
  \end{center}
\end{table}

To that end we have divided the stars in two classes (``dwarfs'' and
``supergiants'', following their luminosity class
classification\footnote{For the one giant in our sample, \cygob~\#4,
we have adopted the line weighting scheme for dwarfs.}), and three
groups in each class (following spectral types). We have then a total
of six stellar groups, and have assigned the spectral lines different
weights depending on their behaviour in each stellar group. This
behaviour represents the expertise from years of ``by eye'' data
analysis that is being translated to the method. Three different
weights are assigned to each line: high, to lines very reliable for
the analysis; medium, and low. The implementation of these weights
into the fitness definition is given by
\begin{equation}
  \label{eq:fitns_weights}
  F \equiv \left(\sum_i^N w_i \chi_{{\rm red}, i}^2\right)^{-1}~,
\end{equation}
where the parameter $w_i$ corresponds to the weight of a specific
line.

\begin{table*}
  \caption{Results obtained for the investigated early type stars
  using GA optimized spectral fits. The spectra were fitted by
  evolving a population of 72 \fastwind\ models over a course of 150
  generations. Spectroscopic masses \Ms\ are calculated with the
  gravities corrected for centrifugal acceleration $\log g_{\rm
  c}$. Evolutionary masses \Mev\ are from \cite{schaller92}. The error
  bars on the derived parameters are given in Tab.~\ref{tab:errors}
  and are discussed in Sect.~\ref{sec:errors}.}
  \label{tab:fit-results}
  \begin{center}
  \begin{tabular}{lccccccrclcc}
  \hline\\[-9pt] \hline \\[-7pt]
  Star & \teff & \logg & \loggc & \rstar & $\log \lstar$ & \yhe
  & \multicolumn{1}{c}{\vturb} & \mdot & \multicolumn{1}{c}{$\beta$} & \Ms & \Mev\\[2pt]
  & [kK] & [\cmsecsec] & [\cmsecsec] & [\rsun] & [\lsun] &  & \multicolumn{1}{c}{[\kmsec]} & [\msunyr] &
  & [\msun] & [\msun]\\[1pt]
 \hline \\[-9pt]
  \cygob~\#7  & 45.8 & 3.93 & 3.94 & 14.4 & 5.91 & 0.21 & 19.9 & 9.98$\cdot10^{-6}$ & 0.77        & 65.1 & 67.8\\[3.5pt]
  \cygob~\#11 & 36.5 & 3.62 & 3.63 & 22.1 & 5.89 & 0.10 & 19.8 & 7.36$\cdot10^{-6}$ & 1.03        & 75.9 & 55.6\\[3.5pt]
  \cygob~\#8C & 41.8 & 3.73 & 3.74 & 13.3 & 5.69 & 0.13 & 0.5  & 3.37$\cdot10^{-6}$ & 0.85        & 36.0 & 49.2\\[3.5pt]
  \cygob~\#8A & 38.2 & 3.56 & 3.57 & 25.6 & 6.10 & 0.14 & 18.3 & 1.04$\cdot10^{-5}$ & 0.74        & 89.0 & 74.4\\[3.5pt]
  \cygob~\#4  & 34.9 & 3.50 & 3.52 & 13.7 & 5.40 & 0.10 & 18.9 & 8.39$\cdot10^{-7}$ & 1.16        & 22.4 & 32.5\\[3.5pt]
  \cygob~\#10 & 29.7 & 3.23 & 3.24 & 29.9 & 5.79 & 0.08 & 17.0 & 2.63$\cdot10^{-6}$ & 1.05        & 56.0 & 45.9\\[3.5pt]
  \cygob~\#2  & 28.7 & 3.56 & 3.57 & 11.3 & 4.88 & 0.08 & 16.5 & 1.63$\cdot10^{-7}$ & 0.80$^{1)}$ & 17.0 & 18.7\\[3.5pt]
  \HD15629    & 42.0 & 3.81 & 3.82 & 12.6 & 5.64 & 0.10 & 8.6  & 9.28$\cdot10^{-7}$ & 1.18        & 37.8 & 47.4\\[3.5pt]
  \HD217086   & 38.1 & 3.91 & 4.01 & 8.30 & 5.11 & 0.09 & 17.1 & 2.09$\cdot10^{-7}$ & 1.27        & 25.7 & 28.5\\[3.5pt]
  \tenlac\    & 36.0 & 4.03 & 4.03 & 8.27 & 5.01 & 0.09 & 15.5 & 6.06$\cdot10^{-8}$ & 0.80$^{1)}$ & 26.9 & 24.9\\[3.5pt]
  \zoph       & 32.1 & 3.62 & 3.83 & 8.9  & 4.88 & 0.11 & 19.7 & 1.43$\cdot10^{-7}$ & 0.80$^{1)}$ & 19.5 & 20.3\\[3.5pt]
  \tausco     & 31.9 & 4.15 & 4.15 & 5.2  & 4.39 & 0.12 & 10.8 & 6.14$\cdot10^{-8}$ & 0.80$^{1)}$ & 13.7 & 16.0\\[1pt]
  \hline
  \end{tabular}
  \end{center}
$^{1)}$ assumed fixed value
\end{table*}

Table~\ref{tab:data} gives the weights assigned to each line in each
stellar group. We will only briefly comment on the low or medium
weights. \hei\ singlets are assigned a low weight for mid-type stars
because of the singlet differential behaviour found between \fastwind\
and \cmfgen\ \citep{puls05}, while they are very weak for early-type
stars. In these two cases therefore we prefer to rely on the triplet
\hei\,$\lambda$4471 line. To this line, however, a low weight is
assigned at late-type Supergiants because of the above mentioned
dilution effect.

\heii\,$\lambda$4686 is only assigned a medium weight (except for late
type dwarfs), as this line is not always completely consistent with
the mass-loss rates derived from Halpha. \heii\,$\lambda$4200 is
sometimes blended with \ion{N}{iii}\,$\lambda$4200, and sometimes it
is not completely consistent with the rest of the \heii\ lines. \hei\
and \heii\ lines at 4026~\AA\ do overlap, but for both lines we find a
consistent behaviour.

The highest weight is therefore given to the Balmer lines plus the
\heii\,$\lambda$4541 and the \hei/\heii\ 4026 lines, which define the
He ionization balance with \hei\,$\lambda$4471 or the singlet \hei\
lines. Note however that, as discussed above, all lines fit
simultaneously in a satisfactory way for our best fitting models.

\subsection{Fits and comments on the individual analysis}

In the following we will present the fits that were obtained by the
automated method for our sample of 12 early type stars, and comment on
the individual analysis of the objects. Listed in
Tab.~\ref{tab:fit-results} are the values determined for the six free
parameters investigated and quantities derived from these.

\subsubsection{Analysis of the \cygob\ stars}
The \cygob\ objects studied here were previously analysed by
\citet[hereafter \citetalias{herrero02}]{herrero02}. We opted to
reanalyse these stars (to test our method) as these stars have equal
distances and have been analysed in a homogeneous way using (an
earlier version of) the same model atmosphere code. In
Sect.~\ref{sec:comp} we will systematically compare our results with
those obtained by \citetalias{herrero02}. Here, we will incidentally
discuss the agreement if this turns out to be relatively poor or if
the absolute value of a parameter seems unexpected, and we wanted to
test possible causes for the discrepancy.

\begin{figure*}
\centering
  \resizebox{17.5cm}{!}{\includegraphics{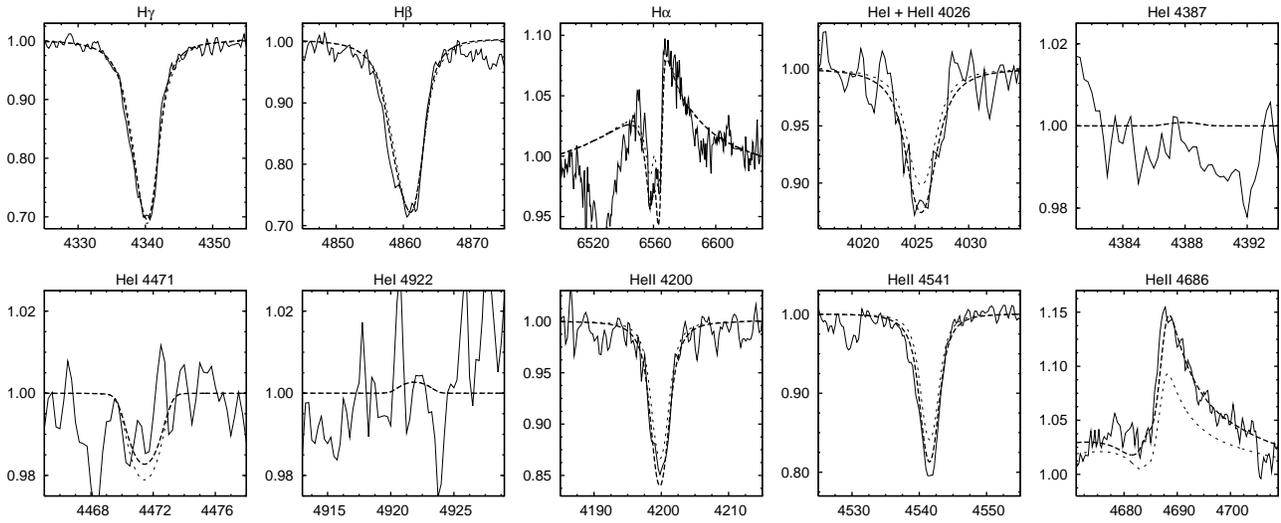}}
  \caption{Comparison of the observed line profiles of \cygob~\#7 with
  the best fit obtained by the automated fitting method (dashed
  lines). Note that the \heii\ line at 6527.1~\AA\ is not included in
  the fit and, therefore, disregarded by the automated
  method. Horizontal axis gives the wavelength in \AA. Vertical axises
  give the continuum normalized flux and are scaled differently for
  each line. In this figure the dotted lines correspond to a fit
  obtained for a helium abundance fixed at 0.1. See text for further
  comments.}
  \label{fig:cob2_7_lines}
\end{figure*}

\paragraph{\cygob~\#7}
The best fit obtained with our automated fitting method for \cygob~\#7
is shown in Fig.~\ref{fig:cob2_7_lines}. For all hydrogen lines
fitted, including \hd\ not shown here, and all \heii\ lines the fits
are of very good quality. Note that given the noise level the fits of
the \hei\ lines are also acceptable.

Interesting to mention is the manner in which the \hei\ and \heii\
blend at 4026~\AA\ is fitted. At first sight, i.e.\ ``by eye'', it
seems that the fit is of poor quality, as the line wings of the
synthetic profile runs through ``features'' which might be attributed
to blends of weak photospheric metal lines. However, the broadest of
these features have a half maximum width of $\sim$70~\kmsec, which is
much smaller than the projected rotational velocity of
105~\kmsec. Consequently, these features are dominated by pure noise.

Compared to the investigation of \citetalias{herrero02} we have
partial agreement between the derived parameters. The mass loss rate,
\teff\ and to a lesser degree $\beta$ agree very well. For \logg\ and
the helium abundance we find, however, large differences. The \logg\
value obtained here is $\sim$0.2~dex larger, which results in a
spectroscopic mass of 65.1~\msun. A value which is in good agreement
with the evolutionary mass of 67.8~\msun.

The helium abundance needed to fit this object is 0.21, which is
considerably lower than the value obtained by \citetalias{herrero02},
who found an abundance ratio of 0.31. This large value still
corresponds to a strong helium surface enrichment. An interesting
question we need to address, is whether this is a real enrichment and
not an artifact that is attributable to a degeneracy effect of \teff\
and \yhe. The latter can be the case, as no \hei\ lines are present in
the optical spectrum of \cygob~\#7. This issue can be resolved with
our fitting method by refitting the spectrum with a {\em helium
abundance fixed} at a lower value than previously obtained. If \teff\
and \yhe\ are truly degenerate this would again yield a good fit,
however for a different \teff.

Shown as dotted lines in Fig.~\ref{fig:cob2_7_lines} are the results
of refitting \cygob~\#7 with a helium abundance fixed at the solar
value. For this lower \yhe\ a \teff\ that is lower by $\sim$2.1~kK was
obtained. This was to be expected as for this temperature regime
\heiii\ is the dominant ionization stage. When consequently \yhe\ is
reduced a reduction of the temperature is required to fit the \heii\
lines. The reduction of \teff\ obtained is the maximum for which still
a good fit of the hydrogen lines is possible and the \hei\ lines do
not become too strong. More importantly, in
Fig.~\ref{fig:cob2_7_lines} it is shown that even with this large
reduction of \teff\ the \heii\ lines cannot be fitted. This implies
that \teff\ and \yhe\ are not degenerate and the obtained helium
enrichment is real.

\begin{figure*}
\centering
  \resizebox{17.5cm}{!}{\includegraphics{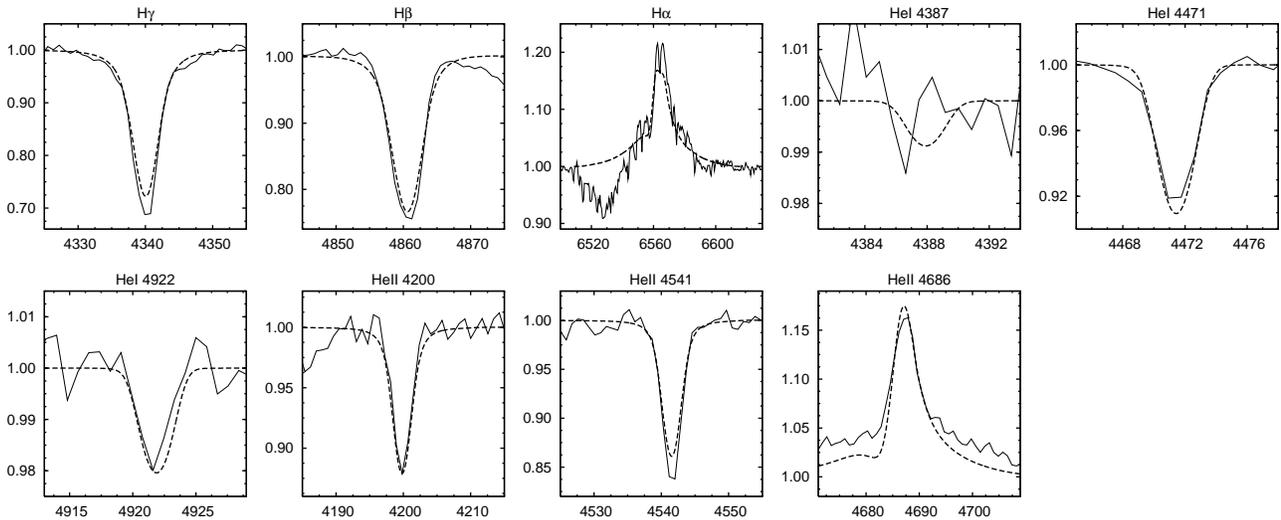}}
  \caption{Same as Fig.~\ref{fig:cob2_7_lines}, however for
  \cygob~\#11}
  \label{fig:cob2_11_lines}
\end{figure*}

\begin{figure*}
\centering
  \resizebox{17.5cm}{!}{\includegraphics{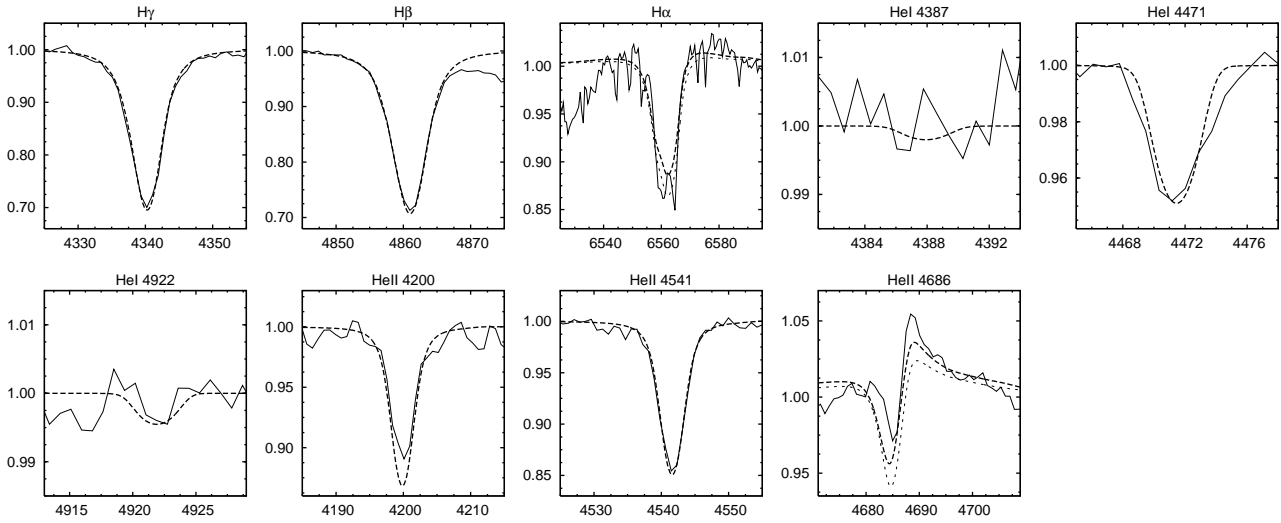}}
  \caption{Same as Fig.~\ref{fig:cob2_7_lines}, however for
  \cygob~\#8C. Shown with a dotted line for \ha\ and
  \heii~$\lambda$4686 are the line profiles of a model with a 0.05~dex
  lower \mdot, which ``by eye'' fits the core of \ha. See text for
  further comments.}
  \label{fig:cob2_8C_lines}
\end{figure*}

\begin{figure*}
\centering
  \resizebox{17.5cm}{!}{\includegraphics{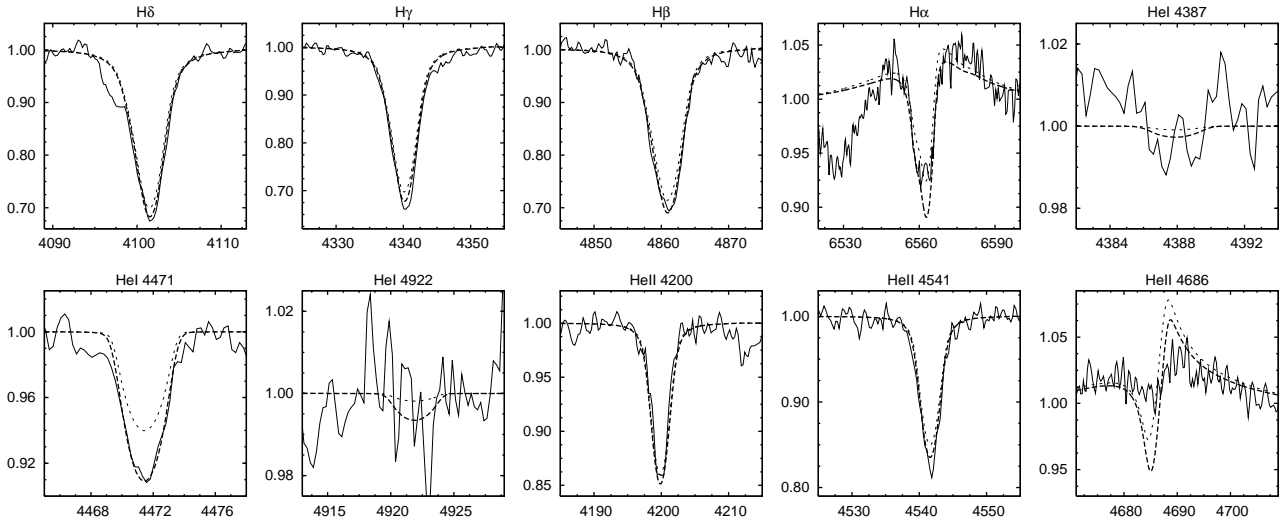}}
  \caption{Same as Fig.~\ref{fig:cob2_7_lines}, however for
  \cygob~\#8A. The dotted lines correspond to a model with a \mdot\
  higher by 0.04 dex. This mass loss rate was obtained by fitting the
  best fit model, found by the automated method, ``by eye'' to the
  \ha\ core.  Even though the fit obtained with the higher \mdot\
  results in a fit of \ha\ which is more pleasing to the eye in the
  line core, this higher mass loss rate does not describe this object
  the best. This can be seen best from the reduced fit quality of the
  other hydrogen Balmer lines and the severe mismatch of
  \hei~$\lambda$4471. See text for further comments.}
  \label{fig:cob2_8A_lines}
\end{figure*}

\paragraph{\cygob~\#11}
Figure \ref{fig:cob2_11_lines} shows the fit to \cygob~\#11. In
general all lines are reproduced correctly. There is a slight
under prediction of the cores of \hg\ and \heii~$\lambda$4541, a
problem that was also pointed out by \cite{herrero92} and
\citetalias{herrero02}. Possibly this is due to too much filling in of
the predicted profiles by wind emission. Part of the
\heii~$\lambda$4541 discrepancy might be related to problems in the
theoretical broadening functions \cite[see][]{repolust05}.

The parameters obtained for this object, with exception of \mdot, are
in agreement with the parameters derived by \citetalias{herrero02}.
With our automated method a mass loss rate lower by $\sim$0.1~dex was
obtained. Note that due to this lower value the behaviour of this
object in terms of its modified wind momentum
(cf. Sect.~\ref{sec:wind-param}) is in better accord with that of the
bulk of the stars investigated in this paper.

\begin{figure*}
\centering
  \resizebox{17.5cm}{!}{\includegraphics{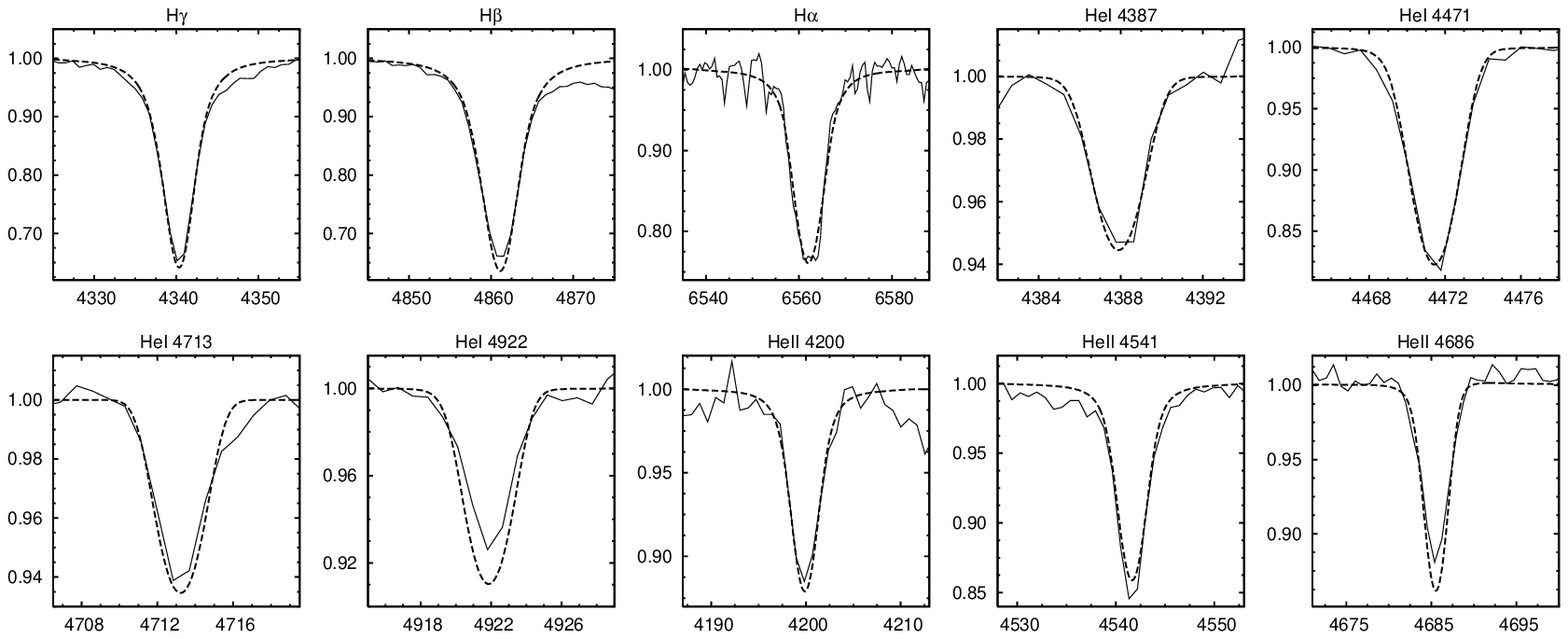}}
  \caption{Same as Fig.~\ref{fig:cob2_7_lines}, however for
  \cygob~\#4.}
  \label{fig:cob2_4_lines}
\end{figure*}

\begin{figure*}
\centering 
  \resizebox{17.5cm}{!}{\includegraphics{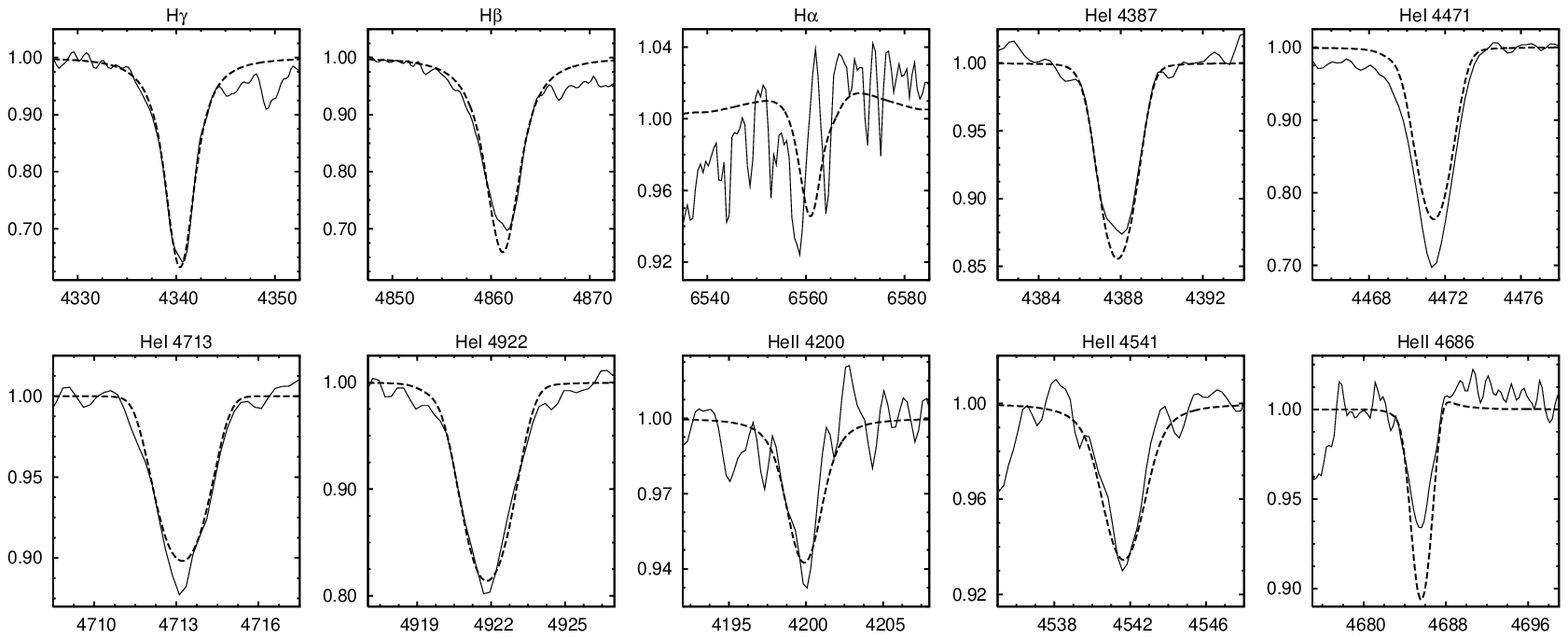}}
  \caption{Same as Fig.~\ref{fig:cob2_7_lines}, however for
  \cygob~\#10.  The emission feature in the core of \ha\ was not
  included in the fit. A subsequent test which did include this
  feature in the fit yielded the same parameters except for a small
  increase of \mdot\ with 0.04~dex.}
  \label{fig:cob2_10_lines}
\end{figure*}

\paragraph{\cygob~\#8C}
The best fit for \cygob~\#8C is shown in Fig.~\ref{fig:cob2_8C_lines}.
Again, with exception of the mass loss rate, the parameters we obtain
for this object are in good agreement with the findings of
\citetalias{herrero02}. We do find a small helium abundance
enhancement, whereas \citetalias{herrero02} found a solar value.

To fit the P~Cygni type profile of \heii\ at 4686~\AA, the automated
method used a \mdot\ which, compared to these authors, was higher by
approximately 0.15 dex. This higher value for the mass loss rate
results in a \ha\ profile which, at first sight, looks to be filled in
too much by wind emission. To assess whether this could correspond to
a significant overestimation of the mass loss rate, we lowered \mdot\
in the best fit model by hand until the core of \ha\ was fitted. In
Fig.~\ref{fig:cob2_8C_lines} the resulting line profiles are shown as
a dotted line for \ha\ and \heii~$\lambda$4686, which for this fit are
the lines which visibly reacted to the change in mass loss rate. To
obtain this fit ``by eye'' of the \ha\ core, a reduction of \mdot\
with merely 0.05 dex was required, showing that the mass loss rate was
not overestimated by the automated method. Note that for this lower
mass loss rate the fit of the \heii~$\lambda$4686 becomes
significantly poorer.

\paragraph{\cygob~\#8A}
\cite{debecker04} report this to be a O6\,I and O5.5\,III binary
system, therefore the derived parameters, in particular the
spectroscopically determined mass, should be taken with care. However,
as this paper also aims to test automated fitting we did pursue the
comparison of this object with \citetalias{herrero02}, who also
treated the system assuming it to be a single star.

We obtained a good fit for all lines except for the problematic
\heii~$\lambda$4686 line. The best fit is shown in
Fig.~\ref{fig:cob2_8A_lines}. Again the \ha\ core is not fitted
perfectly. To determine how significant this small discrepancy is, we
fitted the \ha\ core in a similar manner as for \cygob~\#8C. To obtain
a good fit ``by eye'' we find that \mdot\ has to be increased by 0.04
dex, indicating the extreme sensitivity of \ha\ to \mdot\ in this
regime. The profiles corresponding to the increased mass loss rate
model are shown in Fig.~\ref{fig:cob2_8A_lines} as dotted lines. It is
clear that not only the ``classical'' wind lines react strongly to
\mdot. All synthetic hydrogen Balmer line profiles show significant
filling in due to wind emission for an increased mass loss,
deteriorating the fit quality. Also the \hei~$\lambda$4471 line shows
a decrease in core strength which is comparable to the decrease in the
\ha\ core. This reconfirms that in order to self-consistently
determine the mass loss rate all lines need to be fitted
simultaneously. Therefore, a small discrepancy in the \ha\ core
between the observed and synthetic line profile should not be
considered a decisive reason to reject a fit.

Except for \yhe\ the obtained parameters agree with the results of
\citetalias{herrero02} within the errors given by these
authors. Similar to \cygob~\#8C we find a small helium enhancement.

\paragraph{\cygob~\#4}
The final fit to the spectrum of \cygob~\#4 is presented in
Fig.~\ref{fig:cob2_4_lines}. We obtained good fits for all lines, with
exception of the helium singlet at 4922~\AA, for which the core is
predicted too strong. However, recall that for this spectral type we
assigned a relatively low weight to this line, for reasons explained
in Sect.~\ref{sec:line-scheme}.

The parameters obtained from the fit agree well with the values of
\citetalias{herrero02}, with exception of $\beta$, for which we find a
value higher by $\sim$0.2. Note that \citetalias{herrero02} used a
fixed value for $\beta$ to obtain their fit, whereas in this case the
automated method self-consistently derived this parameter.

The spectroscopic mass implied by the obtained \logg\ value is
significantly smaller than the evolutionary mass of
\cygob~\#4. However, within the error bars (Sect.~\ref{sec:errors})
the two masses agree with each other.

\begin{figure*}
\centering
  \resizebox{17.5cm}{!}{\includegraphics{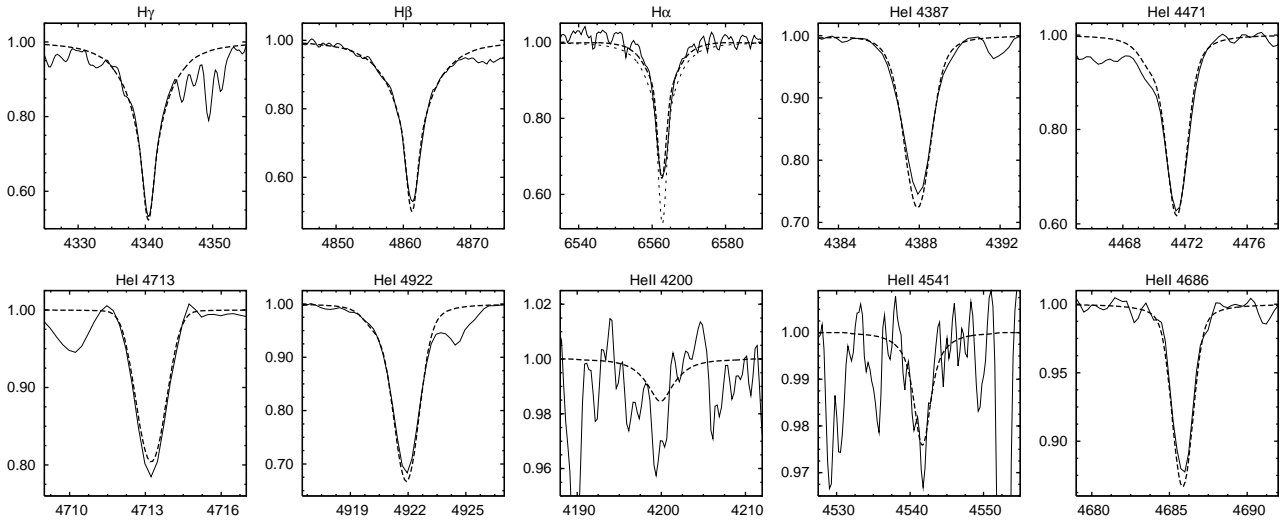}}
  \caption{Same as Fig.~\ref{fig:cob2_7_lines}, however for
  \cygob~\#2. Shown with dotted lines for \ha\ is the line profile of
  the best fit model with a \mdot\ lower by a factor of 3. See text
  for further comments.}
  \label{fig:cob2_2_lines}
\end{figure*}

\paragraph{\cygob~\#10}
In the final fit for this object, shown in
Fig.~\ref{fig:cob2_10_lines}, there are two problematic lines.  First,
for the \heii~$\lambda$4686 line the core is predicted too
strong. Even though compared to \citetalias{herrero02} the situation
has improved considerably, the current version of \fastwind\ still has
difficulties predicting this line. Second, the predicted
\hei~$\lambda$4471 is too weak. Possibly this is connected to the
generalized dilution effect, for which we refer to \cite{repolust04}
for a recent discussion.

In Fig.~\ref{fig:cob2_10_lines} we also see that the \ha\ core of
\cygob~\#10 exhibits an emission feature. For this analysis we assumed
that is was nebular and, consequently, excluded it from the fit. To
test what the effect would be if this assumption was incorrect, a fit
was made with this feature included in the profile. It turned out that
the only parameter which was affected in this test was \mdot, which
showed a small increase of 0.04~dex.

\paragraph{\cygob~\#2}
For \cygob~\#2 the automated method could not self-consistently
determine $\beta$. Therefore, we fixed its value at a theoretically
predicted $\beta=0.8$ \citep[cf.][]{pauldrach86}. In
Fig.~\ref{fig:cob2_2_lines} the best fit is shown. We obtained good
fits for all lines. However, in the case of \hei~$\lambda$4471 we do
see a small under prediction of the forbidden component at 4469~\AA,
which is likely related to incorrect line-broadening functions.

For \logg\ and \mdot\ the obtained fit parameters differ considerably
from the findings of \citetalias{herrero02}. We first focus on mass
loss for which we obtain the relatively low rate of $1.63 \times
10^{-7}$~\msunyr\, with an error bar in the logarithm of this value of
$-0.15$ and $+0.12$ dex (see Tab.~\ref{tab:errors}), given the quoted
value of $\beta$. Our \mdot\ value is approximately a factor two
higher than the mass loss rate obtained by \citetalias{herrero02}.
These authors noted that it was not possible to well constrain the
mass loss rate of such a weak wind. Given the relatively modest errors
indicated by our automated fitting method, we conclude that at least
in principle our technique allows to determine mass loss rates of
winds as weak as that of \cygob~\#2. We have added the phrase ``in
principle'' as it assumes the notion of beta and a very reliable
continuum normalization, which is in this is case different from the
one used by \citetalias{herrero02} for the \heii~4541, \heii~4686 and
\ha\ lines. If this can not be assured, then systematic errors may
dominate over the characteristic fitting error and the mass loss may
be much less well constrained. Assuming the continuum location to be
reliable, we illustrate the sensitivity of the spectrum to mass loss
rates of $\sim$$10^{-7}$ \msunyr\ by reducing the mass loss by a
factor of three. The \ha\ profile of this reduced mass loss model is
shown in Fig.~\ref{fig:cob2_2_lines} as a dotted line. Comparison of
these two cases shows that for winds of order $10^{-7}$ \msunyr\ the
line still contain considerable \mdot\ information. Interestingly, if
we would not take into consideration the line core of \ha\ in our
fitting method, we still recover the quoted mass loss to within three
percent. We note that for our higher mass loss this object appears to
behave well in the wind momentum luminosity relation (see
Sect.~\ref{sec:wind-param}), whereas \citetalias{herrero02} signal a
discrepancy when using their estimated \mdot\ value.

The \logg\ value obtained in this study is 0.36~dex larger than the
value obtained ``by eye'' by \citetalias{herrero02}. Judging from the
very good fits obtained, there is no indication that the automated fit
overestimated the gravity. The spectroscopic mass of 17.0~\msun\
implied by the larger \logg\ value, is also in good agreement with the
evolutionary mass of \cygob~\#2, which is 18.7~\msun.

\subsubsection{Analysis of well studied dwarf OB-stars}

We have also reanalysed five well known and well studied dwarf
OB-stars, sampling the range of O spectral sub-types, in order to
probe a part of parameter space that is not well covered by the
\cygob\ stars. \HD217086 and \zoph, for instance, are fast rotators
with \vsini\ = 350 and 400 \kmsec\ respectively (see also
Tab.~\ref{tab:data}). \tenlac\ is a slow rotator, and \tausco\ is a
very slow rotator. The latter two stars also feature very low mass
loss rates, moreover, the actual \mdot\ values of these stars are much
debated \cite[see][]{martins04}. \HD15629 is selected because it
appears relatively normal.

\begin{figure*}
\centering
  \resizebox{17.5cm}{!}{\includegraphics{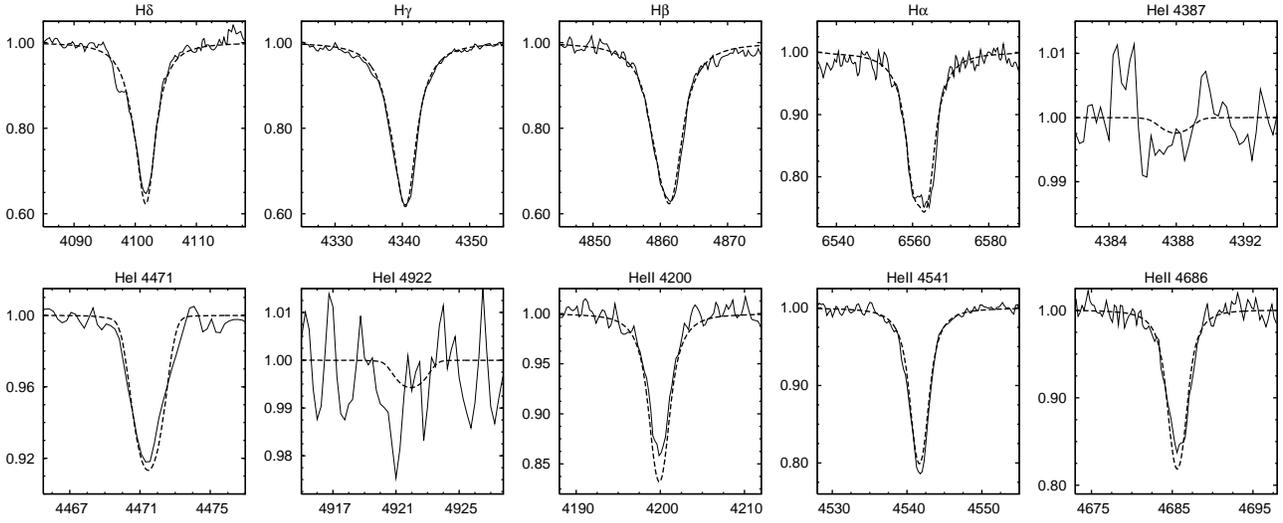}}
  \caption{Same as Fig.~\ref{fig:cob2_7_lines}, however for \HD15629.}
  \label{fig:hd15629_lines}
\end{figure*}

\begin{figure*}
\centering
  \resizebox{17.5cm}{!}{\includegraphics{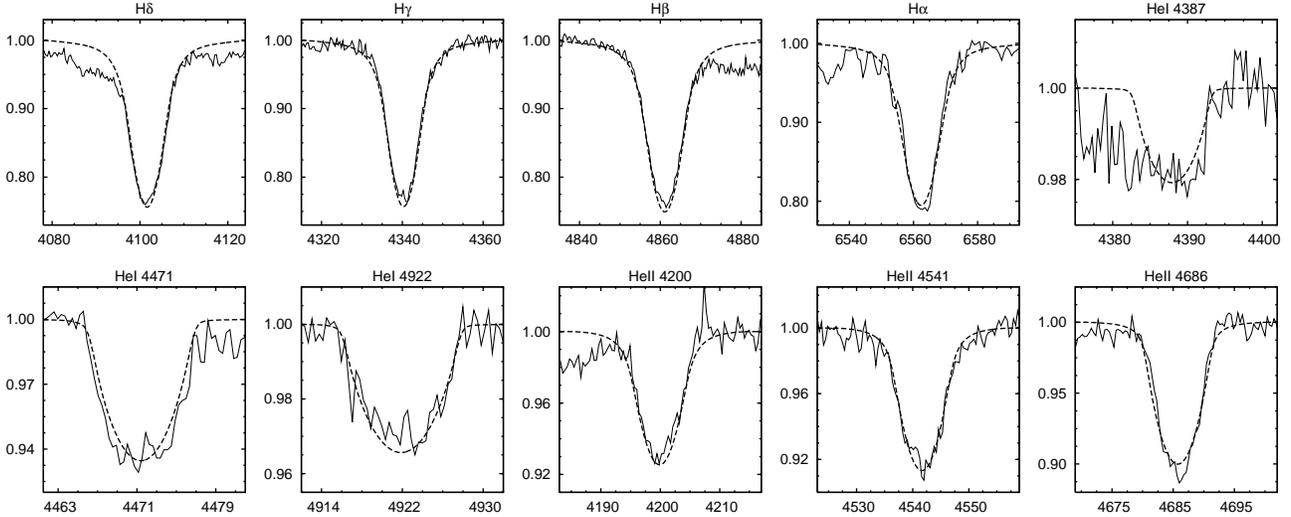}}
  \caption{Same as Fig.~\ref{fig:cob2_7_lines}, however for \HD217086.}
  \label{fig:hd217086_lines}
\end{figure*}

\paragraph{\HD15629}
Apart from a slight over-prediction of the core strength in
\heii~$\lambda$4200, a very good fit was obtained for this object. The
final fit is presented in Fig.~\ref{fig:hd15629_lines}. This object
has recently been studied by \citet[hereafter
\citetalias{repolust04}]{repolust04}. Compared to the parameters
obtained by these authors, we find good agreement except for \teff,
\mdot\ and $\beta$. Note that in contrast to this study we do not find
a helium deficiency. However, the difference of 0.02 with respect to
the solar value obtained here is within the error quoted by
\citetalias{repolust04}.

The difference in wind parameters can be explained by the value of
$\beta=0.8$ assumed by \citetalias{repolust04}. Our self-consistently
derived value for $\beta = 1.18$. As the effect of $\beta$ on the
spectrum is connected to the mass loss rate through the velocity law
and the continuity equation, the lower \mdot\ obtained with the
automated method is explained.

The 1.5~kK increase of \teff\ compared to \citetalias{repolust04} can
be attributed to the improved fit quality and the increase in \logg\
of 0.1~dex. An increase in \logg\ implies an increase in electron
density, resulting in an increase in the recombination rate. The
strength of both the \hei\ and \heii\ lines depend on this rate, as
the involved levels are mainly populated through recombination.
Consequently, as \heiii\ is the dominant ionization stage in the
atmosphere of \HD15629 the strength of the \hei\ and \heii\ lines will
increase when the recombination rate increases. To compensate for this
increase in line strength an increase in \teff, decreasing the
ionization fractions of \hei\ and \heii, is necessary.

\paragraph{\HD217086}
With a projected rotational velocity of 350~\kmsec\ this object can be
considered to be a fast rotator, and our analysis of this object will
show how well the automated method can handle large \vsini. In
Fig.~\ref{fig:hd217086_lines} the best fit obtained with our method is
presented. We find that the large projected rotational velocity does
not pose any problem for the method, i.e.\ the fit quality of all the
lines fitted is very good.

With respect to the obtained parameters, again, these can be compared
to the work of \citetalias{repolust04}. In this comparison we find
considerable differences for \teff\ and \logg\ and a small difference
for \yhe. The effective temperature found by the automated method is
2.1~kK higher. This is a significant increase, but when the \logg\
value obtained here is considered, this can be explained in a similar
manner as the \teff\ increase of \HD15629.

\begin{figure*}
\centering
  \resizebox{17.5cm}{!}{\includegraphics{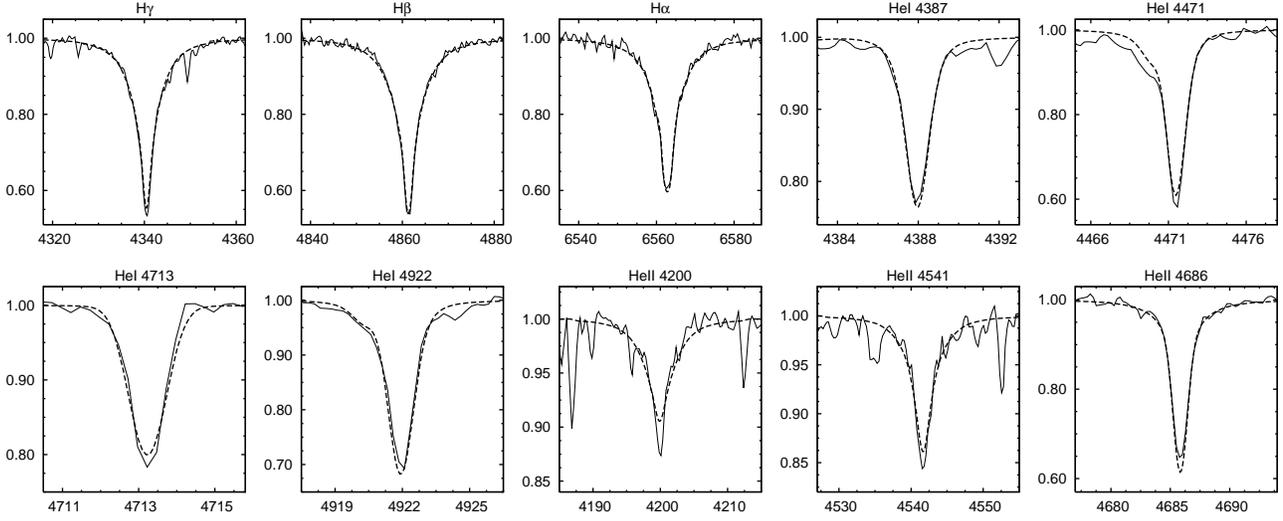}}
  \caption{Same as Fig.~\ref{fig:cob2_7_lines}, however for \tenlac.}
  \label{fig:tenlac_lines}
\end{figure*}

\begin{figure*}
\centering
  \resizebox{17.5cm}{!}{\includegraphics{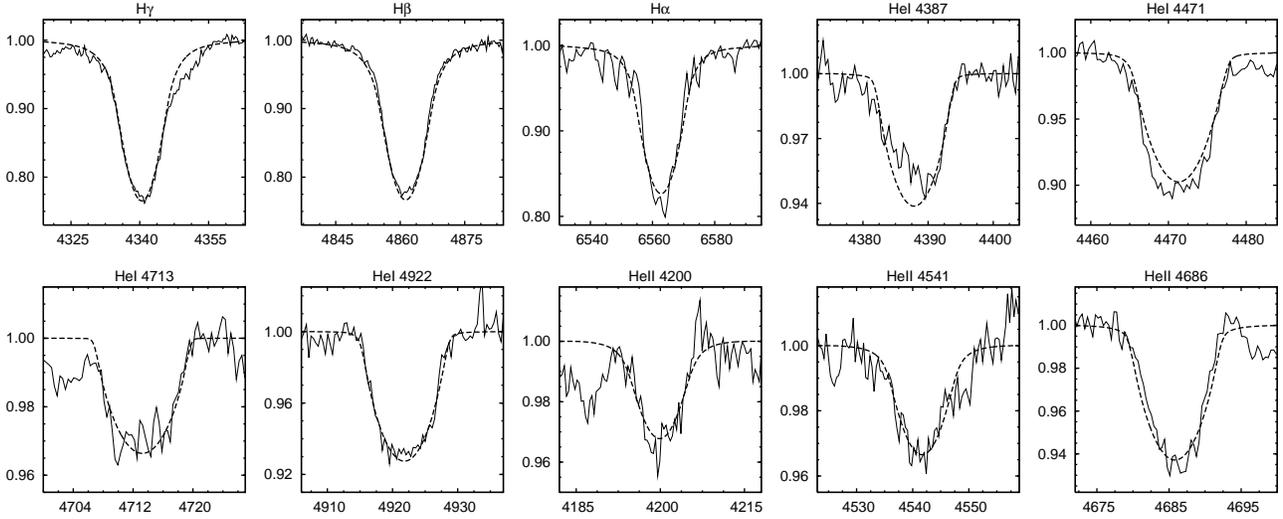}}
  \caption{Same as Fig.~\ref{fig:cob2_7_lines}, however for
  \zoph.}
  \label{fig:zoph_lines}
\end{figure*}

The best fit is obtained with a \logg\ value that is 0.29~dex higher
than the value from \citetalias{repolust04}. Judging from the line
profiles in Fig.~\ref{fig:hd217086_lines} there is no evidence for an
overestimation of \logg. This higher \logg\ removes the discrepancy
with the calibration of \cite{markova04} found by
\citetalias{repolust04} (see Fig.~17 in \citetalias{repolust04}). We
also note that, similar to \cygob~\#2, the increased \logg\ implies a
spectroscopic mass which agrees well with the evolutionary mass of
\HD217086 (cf.\ Tab.~\ref{tab:fit-results}). This is not the case for
the value determined by \citetalias{repolust04}, which points to a
clear discrepancy.

The considerable helium abundance enhancement found by
\citetalias{repolust04} is not reproduced by the automated
method. Even though this object is a rapid rotator, our fit indicates
a normal, i.e.\ solar, helium abundance.

\paragraph{\tenlac}
Like in the case of \cygob~\#2, and for the remaining objects, the
wind is too weak to self-consistently determine $\beta$. Therefore,
again a value of $\beta=0.8$ was assumed.

The photospheric parameters obtained for \tenlac\ agree very well
with the results of \citetalias{herrero02}. The best fit to the
observed spectrum is shown in Fig.~\ref{fig:tenlac_lines}. Whereas
\citetalias{herrero02} find that the mass loss rate cannot be
constrained and only an upper limit of $10^{-8}$ \msunyr\ is found,
the automated method was able to self-consistently determine \mdot\ at
$6 \times 10^{-8}$ \msunyr, though with large error bars (see
Tab.~\ref{tab:errors}). Our error bar indicates that \mdot\ may be an
order of magnitude lower, i.e.\ it may still be consistent with the
\citetalias{herrero02} result.

\begin{figure*}
\centering
  \resizebox{17.5cm}{!}{\includegraphics{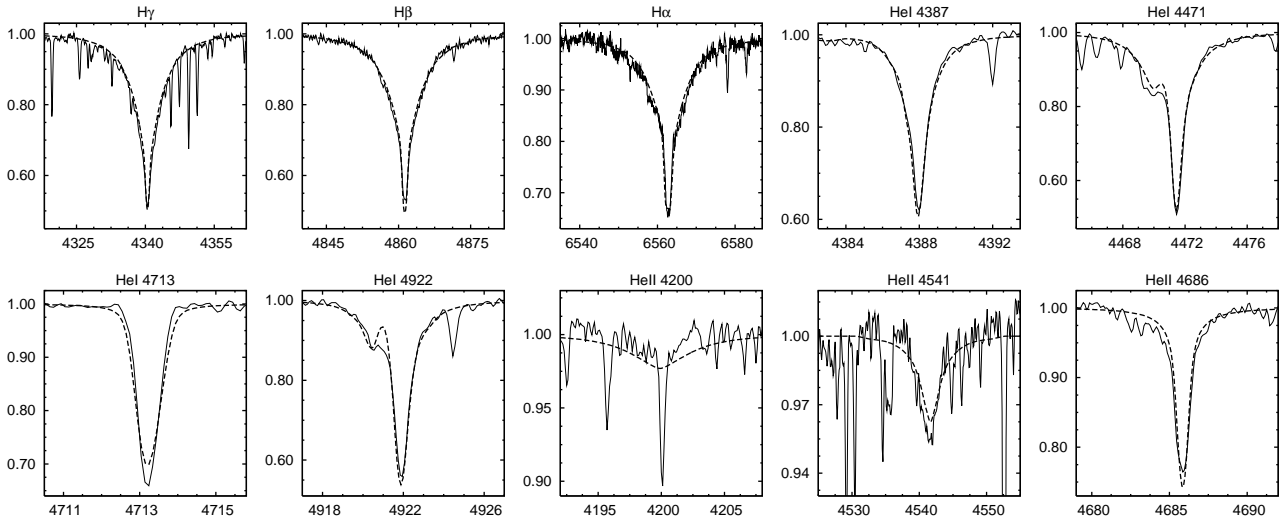}}
  \caption{Same as Fig.~\ref{fig:cob2_7_lines}, however for \tausco.}
  \label{fig:tausco_lines}
\end{figure*}

Various other authors have determined the mass loss rate of \tenlac\
using different methods. These determinations range from up to
$2\times 10^{-7}$ \citep{howarth89} down to $2\times 10^{-9}~\msunyr$
\citep{martins04}. Consequently, compared to these independent
determinations no conclusive answer can be given to the question
whether the \mdot\ derived from the optical spectrum is correct. We
conclude that the mass loss rate of \tenlac\ is anomalously low when
placed into context with the other dwarfs stars studied here. For
instance, the dwarfs \zoph\ and \HD217086, which have luminosities
that are, respectively, lower and higher by $\sim$0.1~dex, both
exhibit a mass loss rate higher by several factors. In
Sect.~\ref{sec:wind-param} we will discuss this further in terms of
the wind-momentum luminosity relation.

\paragraph{\zoph}
The large \vsini\ of 400~\kmsec\ was not a problem to obtain a good
fit. In Fig.~\ref{fig:zoph_lines} the best fit for \zoph\ is
presented.  With the exception of the helium abundance, the comparison
with the results of \citetalias{repolust04} yields very good
agreement. Note that the mass loss rate obtained by these authors is
an upper limit, whereas in this study \mdot\ could be derived
self-consistently. With respect to \yhe\ we do not find any evidence
for a significant overabundance of helium, in agreement with
\cite{villamariz05}.

\paragraph{\tausco}
The best fit for \tausco\ is presented in Fig.~\ref{fig:tausco_lines}.
All lines, including \hd, which is not shown here, are reproduced
accurately. The photospheric parameters we obtained can be compared to
the work of \cite{schonberner88} and \cite{kilian91} who both studied
\tausco\ using plane parallel models. \citeauthor{kilian91} found
\teff=31.7~kK and \logg=4.25, whereas \citeauthor{schonberner88}
obtained \teff=33.0~kK and \logg=4.15. The difference in \teff\
between the two studies is explained by the fact that in the latter
analysis no line blanketing was included in the models. Therefore, we
prefer to compare our \teff\ to the former investigation, which agree
very well. In terms of the gravity we find good agreement with the
second study. The value obtained by \citeauthor{kilian91} seems rather
high. Given the almost perfect agreement between the synthetic line
profiles and the observations in Fig.~\ref{fig:tausco_lines}, the
reason for this discrepancy is unclear. On a side note, more recently
\cite{repolust05} analysed the infrared spectrum of this object. Their
findings do confirm our lower value, but could not reproduce the
enhanced helium abundance we find, due to a lack of observed infrared
\heii\ lines.

In the recent literature the mass loss rate usually adopted for
\tausco\ is $9\times10^{-9}$~\msunyr, which is considerably smaller
than the $6.14\times10^{-8}$~\msunyr\ obtained in this study. However,
the former mass loss rate is an average value determined by
\cite{dejager88}, based on the mass loss rates independently found by
\cite{gathier81} and \cite{hamann81a}. Based on the UV resonance
lines, these two studies, respectively, determined \mdot\ to be
$7.4\times10^{-8}$ and $1.3\times10^{-9}$~\msunyr. So, they differ by
more than a factor of 50. The mass loss rate obtained with the
automated method is in reasonable agreement with that obtained by
\citeauthor{gathier81}. Our higher value is also supported by the
study of the infrared spectrum of \tausco\ by \cite{repolust05} who
find $\mdot \simeq 2\times10^{-8}~\msunyr$. Detailed fitting of \bra\
will likely clarify this issue.

\begin{figure*}
  \centering
  \resizebox{17cm}{!}{\includegraphics{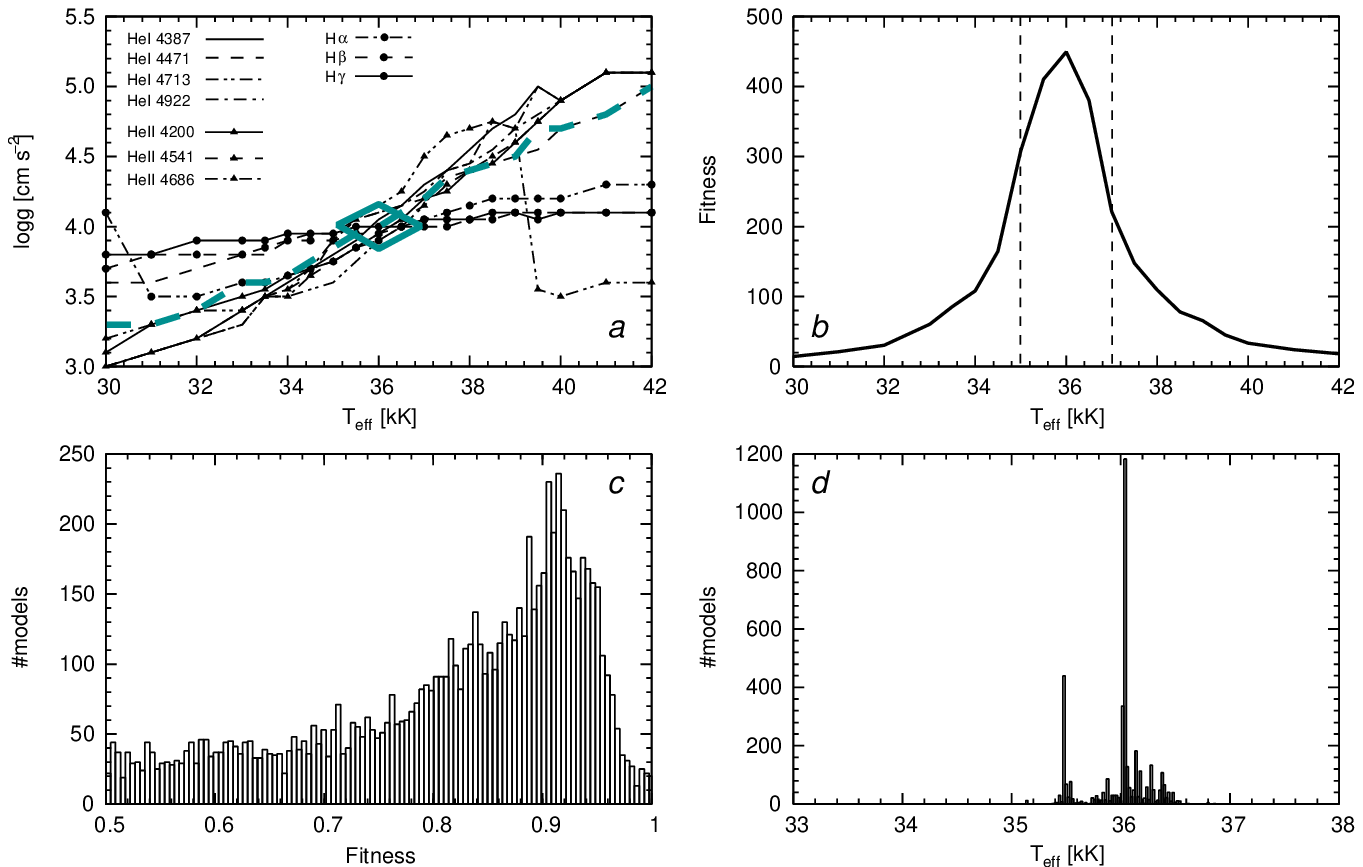}}
  \caption{{\it Panel a:} fit diagram of \teff\ and \logg\ for
  \tenlac. {\it Panel b:} Fitness as a function of \teff\ for $\logg\
  = 4.0$. {\it Panel c:} Fitness distribution of the models calculated
  during the fitting run of the automated method. {\it Panel d:}
  Distribution of \teff\ in the models located within the global
  optimum. The maximum variation of \teff\ within the global optimum,
  which corresponds to the error estimate of this parameter, is
  $\sim$900~K.}
  \label{fig:error-an}
\end{figure*}

\section{Error analysis}
\label{sec:errors}

Here we will introduce our method of estimating errors on the
parameters derived with the automated method. This method is based on
properties of the distribution of the fitnesses of the models in
parameter space, which may seem conceptually different from classical
approaches of defining error bars (and in a sense it is). However, we
will demonstrate for the case of 10\,Lac that our error definition is
very comparable to what is routinely done in fit diagram approaches.

\subsection{Fit diagrams}
In a fit diagram method the error bar on \teff\ and \logg\ is derived
by investigating the simultaneous behaviour of these two parameters.
In panel {\it a} of Fig.~\ref{fig:error-an} the fit diagram of 10\,Lac
is presented adopting for all other parameters (save \teff\ and \logg)
the best fit values obtained in our automated fitting. This diagram
was constructed by calculating a grid of \fastwind\ models in the
\teff-\logg\ plane, and evaluating for every line for every \teff\
which model, i.e.\ \logg, fits this line the best. The location where
the resulting fit curves intersect, corresponds to the best fit. This
best fit yields \teff\ = 36\,000~K and \logg\ = 4.0. Note that this
result was obtained without the use of our automated method. The error
can now be estimated by estimating the dispersion of the fit curves
around this location. In panel {\it a} of Fig.~\ref{fig:error-an} this
is indicated by a box around the best fit location. The corresponding
error estimates are 1000~K in \teff\ and 0.1~dex in \logg.

The method described above cannot be applied to our automated fitting
method due to two reasons. First, as we have defined the fit quality
according to Eq.~(\ref{eq:fitns}), this definition of fitness
compresses the fit curves of all individual lines in the fit diagram
to a single curve. In Fig.~\ref{fig:error-an} this curve is shown as a
thick dashed line. Although the curve runs through the best fit point,
no information about the dispersion of the solutions around this point
can be derived from it. The second reason lies in the multidimensional
character of the problem of line fitting. If one would want to
properly estimate the error taking this multidimensionality in to
account, a fit diagram should be constructed with a dimension equal to
the number of free parameters evaluated. In case of our fits this
translates to the construction of a six dimensional fit diagram.

\subsection{Optimum width based error estimates}
Even though we have argued that fit diagrams cannot be used with our
fitting method, it is possible to construct an error estimate which is
analogous to the use of these diagrams and does take the
multidimensionality of the problem into account. This can be done by
first realizing that the error box shown in Fig.~\ref{fig:error-an}
essentially is a {\em measure of the width of the optimum in parameter
space}, i.e.\ it defines the region in which models are located which
approximately have {\em the same fit quality}. This is illustrated in
panel {\it b}. There we show the one-dimensional fitness function in
the \teff-\logg\ plane for $\logg=4.0$. Indicated with dashed lines is
the error in \teff\ estimated using the fit diagram of
\tenlac. Confined between these lines is the region which corresponds
to the optimum as defined by the error box in panel {\it
a}. Consequently, the difference between maximum and minimum fitness
in this region defines the width of the optimum. Returning to the
general case, we can now invert the reasoning and state that the error
estimate for a given parameter is equal to the maximum variation of
this parameter in the group of best fitting models, i.e.\ the models
located within the error box. Consequently, in the automated fitting
method by defining a group of best fitting models, the {\em error
estimates for all free parameters} can be determined.

\begin{table*}
  \caption{Error estimates for fit parameters obtained using the
  automated fitting method and parameters derived from these.  Denoted
  by {\sc nd} are errors in \vturb\ that reach up to the maximum
  allowed value of \vturb\ and, therefore, are formally not
  defined. Uncertainties in the fit parameters result from the optimum
  width based error estimates method. See text for details and
  discussion.}
  \label{tab:errors}
  \begin{center}
  \begin{tabular}{lccccclclll}
  \hline\\[-9pt] \hline \\[-7pt]
  Star & $\Delta$\teff & $\Delta$\loggc & $\Delta$\rstar & $\Delta \log \lstar$ & $\Delta$\yhe
  & \multicolumn{1}{c}{$\Delta$\vturb} & $\Delta\log$ \mdot & \multicolumn{1}{c}{$\Delta$$\beta$} & \multicolumn{1}{c}{$\Delta$\Ms} & \multicolumn{1}{c}{$\Delta$\Mev}\\[2pt]
  & [kK] & [\cmsecsec] & [\rsun] & [\lsun] &  & \multicolumn{1}{c}{[\kmsec]} & [\msunyr] &
  & \multicolumn{1}{c}{[\msun]} & \multicolumn{1}{c}{[\msun]}\\[1pt]
 \hline \\[-9pt]
  \cygob~\#7  & $^{-1.0}_{+1.5}$ & $^{-0.08}_{+0.06}$ & $\pm$0.7 &
  $\pm$0.07 & $^{-0.02}_{+0.03}$ & \hspace{6pt}$^{-14.9}_{\rm +ND}$ & $^{-0.05}_{+0.03}$ & $^{-0.04}_{+0.09}$ & \hspace{4pt}$^{-15}_{+12}$ & \hspace{4pt}$^{-7 }_{+7}$\\[3.5pt]
  \cygob~\#11 & $^{-0.6}_{+0.4}$ & $^{-0.07}_{+0.13}$ & $\pm$1.1 &
  $\pm$0.05 & $^{-0.01}_{+0.03}$ & \hspace{6pt}$^{-4.0}_{\rm +ND}$  & $^{-0.03}_{+0.06}$ & $^{-0.05}_{+0.02}$ & \hspace{4pt}$^{-15}_{+27}$ & \hspace{4pt}$^{-3 }_{+4}$\\[3.5pt]
  \cygob~\#8C & $^{-1.3}_{+1.1}$ & $^{-0.10}_{+0.14}$ & $\pm$0.7 & $\pm$0.07 & $^{-0.02}_{+0.04}$ & \hspace{6pt}$^{-0.2}_{+10.9}$ & $^{-0.07}_{+0.04}$ & $^{-0.05}_{+0.10}$ & \hspace{4pt}$^{-10}_{+14}$ & \hspace{4pt}$^{-4 }_{+4}$\\[3.5pt]
  \cygob~\#8A & $^{-0.4}_{+1.7}$ & $^{-0.05}_{+0.13}$ & $\pm$1.3 &
  $\pm$0.09 & $^{-0.04}_{+0.04}$ & \hspace{6pt}$^{-17.7}_{\rm +ND}$ & $^{-0.07}_{+0.03}$ & $^{-0.04}_{+0.11}$ & \hspace{4pt}$^{-15}_{+32}$ & \hspace{4pt}$^{-10}_{+8}$\\[3.5pt]
  \cygob~\#4  & $^{-0.3}_{+1.5}$ & $^{-0.04}_{+0.21}$ & $\pm$0.7 &
  $\pm$0.09 & $^{-0.02}_{+0.03}$ & \hspace{6pt}$^{-3.0}_{\rm +ND}$  & $^{-0.10}_{+0.05}$ & $^{-0.05}_{+0.21}$ & \hspace{4pt}$^{-3 }_{+15}$ & \hspace{4pt}$^{-3 }_{+3}$\\[3.5pt]
  \cygob~\#10 & $^{-0.8}_{+1.0}$ & $^{-0.12}_{+0.16}$ & $\pm$1.5 &
  $\pm$0.07 & $^{-0.02}_{+0.03}$ & \hspace{6pt}$^{-7.0}_{\rm +ND}$  & $^{-0.13}_{+0.08}$ & $^{-0.15}_{+0.19}$ & \hspace{4pt}$^{-19}_{+26}$ & \hspace{4pt}$^{-4 }_{+4}$\\[3.5pt]
  \cygob~\#2  & $^{-0.8}_{+1.2}$ & $^{-0.14}_{+0.13}$ & $\pm$0.6 & $\pm$0.08 & $^{-0.01}_{+0.03}$ & \hspace{6pt}$^{-2.3}_{+2.4}$  & $^{-0.15}_{+0.12}$ & -                  & \hspace{4pt}$^{-7 }_{+6 }$ & \hspace{4pt}$^{-1 }_{+2}$\\[3.5pt]
  \HD15629    & $^{-0.3}_{+0.7}$ & $^{-0.05}_{+0.07}$ & $\pm$1.9 & $\pm$0.12 & $^{-0.01}_{+0.03}$ & \hspace{6pt}$^{-8.4}_{+7.6}$  & $^{-0.13}_{+0.10}$ & $^{-0.10}_{+0.27}$ & \hspace{4pt}$^{-13}_{+14}$ & \hspace{4pt}$^{-5 }_{+7}$\\[3.5pt]
  \HD217086   & $^{-0.5}_{+0.9}$ & $^{-0.08}_{+0.07}$ & $\pm$1.2 & $\pm$0.13 & $^{-0.02}_{+0.02}$ & \hspace{6pt}$^{-4.9}_{+2.9}$  & $^{-0.12}_{+0.18}$ & $^{-0.25}_{+0.16}$ & \hspace{4pt}$^{-10}_{+10}$ & \hspace{4pt}$^{-3 }_{+3}$\\[3.5pt]
  \tenlac\    & $^{-0.9}_{+0.8}$ & $^{-0.12}_{+0.13}$ & $\pm$1.7 & $\pm$0.17 & $^{-0.02}_{+0.02}$ & \hspace{6pt}$^{-3.8}_{+4.1}$  & $^{-0.98}_{+0.39}$ & -                  & \hspace{4pt}$^{-16}_{+16}$ & \hspace{4pt}$^{-2 }_{+4}$\\[3.5pt]
  \zoph       & $^{-0.7}_{+0.7}$ & $^{-0.05}_{+0.16}$ & $\pm$1.3 &
  $\pm$0.13 & $^{-0.02}_{+0.04}$ & \hspace{6pt}$^{-6.2}_{\rm +ND}$  & $^{-0.28}_{+0.15}$ & -                  & \hspace{4pt}$^{-7 }_{+11}$ & \hspace{4pt}$^{-2 }_{+2}$\\[3.5pt]
  \tausco     & $^{-0.8}_{+0.5}$ & $^{-0.14}_{+0.09}$ & $\pm$0.5 & $\pm$0.09 & $^{-0.02}_{+0.04}$ & \hspace{6pt}$^{-2.2}_{+2.4}$  & $^{-0.99}_{+0.22}$ & -                  & \hspace{4pt}$^{-6 }_{+4 }$ & \hspace{4pt}$^{-1 }_{+1}$\\[3.5pt]
  \hline
  \end{tabular}
  \end{center}
\end{table*}

We define the group of best fitting models as the group of models that
lie within the global optimum. Put differently, the width of the
global optimum in terms of fitness, defines the group of best fitting
models. Identifying and, consequently, measuring this width is
facilitated by the nature of the GA, i.e.\ selected reproduction,
incorporated in our fitting method. Due to this selected reproduction
the exploration through parameter space results in a mapping of this
space in which regions of high fit quality, i.e.\ the regions around
local optima and the global optimum, are sampled more
intensively. Consequently, if we would rank all models of all
generations calculated during a fitting run according to their
fitness, the resulting distribution will peak around the locations of
the optima. In case of the global optimum the width of this peak,
starting from up to the maximum fitness found, is, analogous to the
width of the error box used in a fit diagram, a direct measure of the
width of the optimum. Consequently, this width depends on the quality
of the data, i.e.\ it will be broader or narrower for, respectively,
low and high signal to noise, and on the degeneracy between the fit
parameters. Therefore, {\em the error estimates of the individual
parameters are equal to the maximum variations of these parameters for
all models contained in the peak corresponding to the global optimum}.

In panel {\it c} of Fig.~\ref{fig:error-an} the distribution of the
models according to their fitness calculated during the fitting run of
\tenlac\ using the automated method is shown. The fitnesses are
normalized with respect to the highest fitness and only the top half
of the distribution is shown. In this distribution two peaks are
clearly distinguishable. The most pronounced peak is located at $F
\approx 0.9$ and corresponds to the region around the global
optimum. A second peak, corresponding to a region around a secondary
optimum, is located at $F \approx 0.83$. To derive the error on the
fit parameters we estimate the total width of the global optimum for
\tenlac\ to be $\sim$0.15\footnote{In general this is not a fixed
number. Considering all programme stars we find the width of the
global optimum to be within the range $\sim$0.1 to $\sim$0.2.}, i.e.\
the range of $F=0.85...1.0$ corresponds to the width of the
optimum. In panel {\it d} of Fig.~\ref{fig:error-an} we show the
resulting distribution of \teff\ of the models within this global
optimum. In this figure we see that the maximum variation, hence the
error estimate, is $\sim$900~K, which is in good agreement with the
value derived using the fit diagram of \tenlac. For \logg\ we also
find an error estimate of $\sim$0.1~dex, which is also very similar to
the value obtained with this diagram. The exact values as well as
error estimates for all fit parameters of all objects are given in
Tab.~\ref{tab:errors}. It is important to note that our error analysis
method also allows for an error estimate of parameters to which the
spectrum does not react strongly. For \tenlac\ this is clearly the
case for the mass loss rate, for which we find large error bars.

\subsection{Derived parameters}
In Tab.~\ref{tab:errors} the errors on the derived parameters were
calculated based on the error estimates of the fit parameters. Here we
will elaborate on their derivation.

The error in the stellar radii is dominated by the uncertainty in the
absolute visual magnitude. In case of the \cygob\ objects we adopt
these to be 0.1\magn\ conform the work of \cite{massey91}. For
\HD15629, \HD217086 and \zoph\ we use the uncertainty as given by
\citetalias{repolust04} of 0.3\magn. The distance to \tenlac\ and
\tausco\ was measured by Hipparcos. Therefore, for these two objects
we adopt the error based on this measurement, which, respectively, is
0.4 and 0.2\magn. Together with the uncertainty in \teff\ the
uncertainty in \rstar\ is calculated according to Eq.~(8) of
\citetalias{repolust04}, where we used the largest absolute
uncertainty in \teff\ for a given object.

To correct the surface gravity for centrifugal forces, a
correction conform \cite{herrero92} was applied to the gravity
determined from the spectral fits. This corrected value is given in
Tab.~\ref{tab:fit-results}. As shown by \citetalias{repolust04} this
correction has a non negligible effect on the error in the resulting
\loggc. Consequently, we used their estimate to calculate the total
error estimate of \loggc\ as given in Tab.~\ref{tab:errors}. Using
this error together with the uncertainty in \rstar\ the resulting
uncertainty in the spectroscopic mass was calculated.

For the calculation of the uncertainty in the stellar luminosity, we
consistently adopted the largest absolute error in \teff. The
resulting $\Delta \log \lstar$ as well as the uncertainty in \teff\
have an effect on the evolutionary mass. We have estimated errors for
this quantity using the error box spanned by $\Delta \log \lstar$ and
$\Delta \log \teff$.

\section{Comparison with previous results}
\label{sec:comp}

\begin{figure}[t]
\centering
  \resizebox{8.8cm}{!}{
    \includegraphics{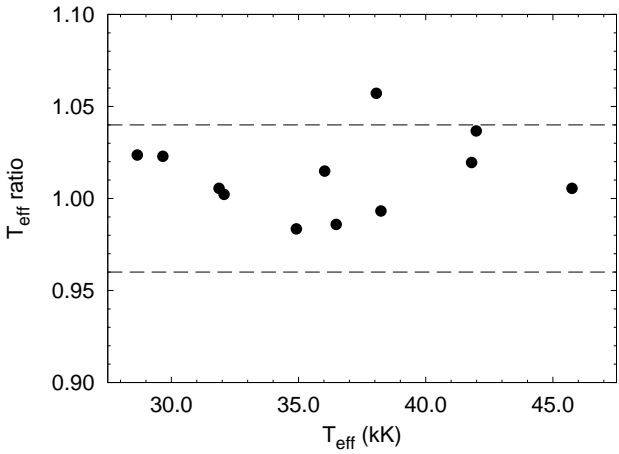}}
  \caption{Comparison of the effective temperatures obtained using
  automated fits (horizontal axis) and ``by eye'' fits. On the
  vertical axis the ratio of automated relative to ``by eye''
  temperature determination is given. The dashed lines correspond to a
  four percent error usually adopted for ``by eye'' determined
  values.}
  \label{fig:teff_comp}
\end{figure}

In this section we will compare the results obtained with our
automated fitting method with those from ``by eye'' fits (relevant
references to the comparison studies are given in the previous
section). This does not constitute a one-to-one comparison of the
automated and ``by eye'' approach as this would require the use of
identical model atmosphere codes as well as the same set of spectra,
moreover, with identical continuum normalization. Potential
differences can therefore not exclusively be attributed to the less
bias sensitive automated fitting method. However, as we have applied
our method to a sizeable sample of early type stars, the automated
nature of it does assure that it is the most homogeneous study to
date, i.e.\ without at least some of the biases involved in
conventional analyses.

\subsection{Effective temperature}

In Fig.~\ref{fig:teff_comp} a comparison of the effective temperatures
determined in this study with \teff\ values obtained with ``by eye''
fits, is presented. Indicated with dashed lines are the four percent
errors usually adopted for ``by eye'' fitted spectra.  With the
exception of the outlier \HD217086 at 38.1~kK, the agreement is very
good and no systematic trend is visible. From this plot we can
conclude that the \teff\ obtained with the automated fit is at least
as reliable as the temperatures determined in the conventional way.

\subsection{Gravities}
\label{sec:gravities}

\begin{figure}[t]
\centering \resizebox{8.8cm}{!}{ \includegraphics{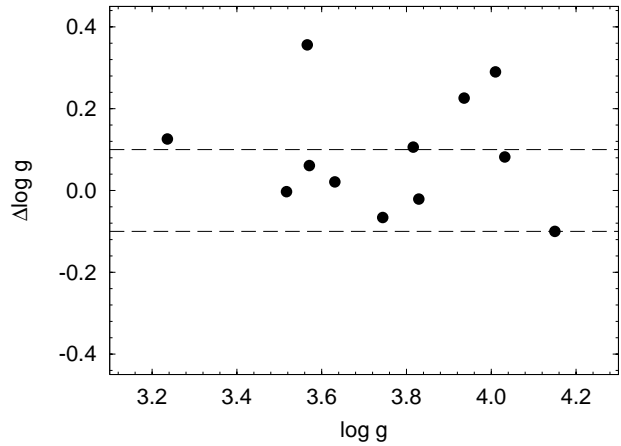}}
  \caption{Gravities obtained with automated fits (horizontal axis)
  are compared to gravities determined from ``by eye'' fits. The
  vertical axis gives the difference of the logarithm of the two
  gravity determinations. Indicated by dashed lines are the 0.1 dex
  errors usually adopted for gravities determined ``by eye''.}
  \label{fig:logg_comp}
\end{figure}

In many cases the gravity obtained with the automated procedure is
significantly higher than the values obtained with the conventional
``by eye'' fitted spectra. This is shown in Fig.~\ref{fig:logg_comp},
where we show as a function of the gravities obtained in this study
the differences with the ``by eye'' determined values. Indicated with
dashed lines in this figure is the 0.1~dex error in \logg\ that is
often assigned to a ``by eye'' fitting of the hydrogen Balmer
line wings. It is important to note that this plot shows that there is
no obvious trend in the differences, i.e.\ there appears no systematic
increase as a function of \logg\ present.

It is clear, however, that there are three outliers for which previous
gravity determinations yield values that are at least 0.2 dex
lower. These are in order of increasing gravity (as determined in this
study): \cygob~\#2, \cygob~\#7 and \HD217086. For all three cases,
previous spectroscopic mass determinations result in values that are
about a factor of two less than the corresponding evolutionary masses.
One reason for these discrepant gravity values can be traced to a
difference between automated and ``by eye'' fitting. In ``by eye''
fitting, it is custom to prohibit the theoretical line flux in the
wings of Balmer lines -- specifically that at the position in the
observed line wing where the profile curvature is maximal -- to be
below that of the observed flux. This constraint has been used by
\citetalias{herrero02} for the \cygob\ stars; for \HD217086, we could
not verify whether this was the case. The automated method does not
apply this constraint. Therefore, as it strives for a maximum fitness,
it tends to fit the curve through the signal noise as much as
possible. This yields a higher gravity.

A second reason is connected to the multidimensional nature of the
optimization problem. ``By eye'' fitting may not find the optimum fit,
as in general it can not simultaneously deal in a sufficiently
adequate way with all the free parameters of the problem.
Consequently, some of the ``by eye'' fitted spectra do not correspond
to the best fit possible. A good example in which this appears to be
the case is \HD217086. With the automated fit we not only obtained a
gravity that is higher by $\sim$0.3~dex, but also an effective
temperature higher by 2.1~kK compared to the results of
\citetalias{repolust04}. Consequently, as the ionization structure of
the atmosphere depends heavily on this temperature, so does the
gravity one obtains from a spectral fit for this temperature. As
\citetalias{repolust04} obtained a gravity for a significantly lower
effective temperature, the gravity obtained from their spectral fit
likely corresponds to the value from a local optimum in parameter
space.

\subsection{Helium abundance and microturbulence}

This analysis is the first in which the helium abundance and the
microturbulent velocity have been treated as {\em continuous} free
parameters. In the studies of \citetalias{herrero02} and
\citetalias{repolust04} only two possible values for the
microturbulent velocity were adopted. For the helium abundance an
initial solar abundance was adopted, which was modified when no
satisfying fit could be obtained for this abundance.  Consequently, a
comparison with these studies as was done for e.g.\ the gravities, is
not possible. Instead we will only discuss whether the obtained values
of these parameters are reasonable and comment on possible
correlations with other parameters.

The helium abundances given in Tab.~\ref{tab:fit-results} show that no
extreme values were needed by the fitting method to obtain a good
fit. An exception to this may be \yhe=0.21 obtained for
\cygob~\#7. However, as discussed earlier this value is still
significantly smaller than the \yhe=0.3 obtained by
\citetalias{herrero02}. With respect to a possible relation between
the helium abundance and other parameters, only a small correlation
between \teff\ and \yhe\ is found for the supergiants. For these
objects it appears (cf.\ Tab.~\ref{tab:fit-results}) that the helium
abundance increases with increasing effective temperature. However, as
we only analysed six supergiants further investigation using a larger
sample needs to be undertaken.

Also for the microturbulent velocities no anomalous values were needed
to fit the spectra. The large error bars in the turbulent velocity
quoted in Tab.~\ref{tab:errors}, especially for the supergiants, show
that the profiles are not very sensitive to this parameter. This is
consistent with the study of \cite{villamariz00} and
\citetalias{repolust04}. The {\sc nd} entries given on some of the
positive errors in the table indicate that they reach up to the
maximum allowed value of \vturb, which is 20 \kmsec. Therefore, they
are formally not defined. The fact that some of the small scale
turbulent velocities are close to this maximum value may indicate that
they represent lower limits, though, again, this likely reflects that
they are poorly constraint.

No correlation of the microturbulence with any of the other parameters
is found. In particular not between \vturb\ and \logg\ and \vturb\ and
\yhe. Various authors have hinted at such a correlation
\citep[e.g.][]{kilian92}. 

\subsection{Wind parameters}
\label{sec:wind-param}

\begin{figure}
\centering
  \resizebox{8.8cm}{!}{
  \includegraphics{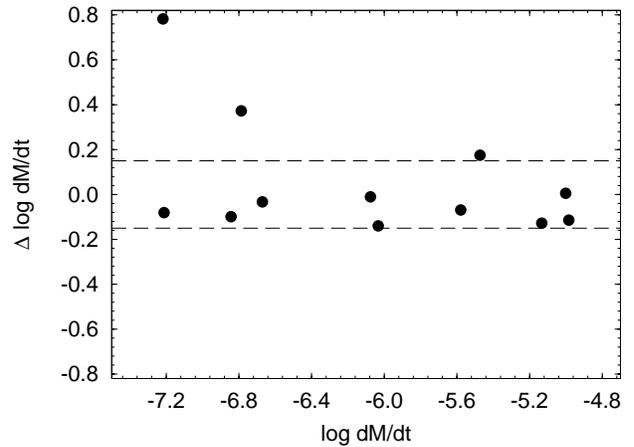}}
  \caption{Difference between mass loss rate obtained by the automated
  method (given by the horizontal) axis and values determined by eye.
  A typical 0.15~dex error is indicated by the dashed lines. The two
  outliers at $\log \mdot \simeq -7.2$ and and $\log \mdot \simeq
  -6.8$, respectively, correspond to \tenlac\ and \cygob~\#2.}
  \label{fig:mdot_comp}
\end{figure}

The straightforward comparison of the mass loss rates obtained with
the automated method with values determined from spectral fits ``by
eye'' is shown in Fig.~\ref{fig:mdot_comp}. With exception of \tausco\
at $\log\mdot = -7.2$, for which the mass loss rate determined by
\cite{gathier81} from UV line fitting serves as a comparison, all mass
loss rates are compared to values determined from \ha\ fitting. For
this comparison we assume an error of 0.15~dex in the ``by eye''
determined values. This uncertainty corresponds to a typical error
obtained from \ha\ fitting and is shown in Fig.~\ref{fig:mdot_comp} as
a set of dashed lines. With exception of \tenlac\ and \tausco, for
which the mass loss rate determination is uncertain, this error is
also comparable to the errors obtained with the automated method.

Two objects show a relative increase in \mdot\ which is much larger
than the typical error. These are \tenlac\ at $\log \mdot \simeq -7.3$
and \cygob~\#2 at $\log \mdot \simeq -6.8$. In the case of the latter
we showed that the increase is due to a more efficient use of wind
information stored in the line profiles by the automated method, which
improves the relation of \cygob~\#2 with respect to the wind-momentum
relation (see Sect.~\ref{sec:wlr}).

With respect to \tenlac\ we already mentioned that a range of more
than two orders of magnitude in mass loss rate has been found in
different studies. Here we have made the comparison with the upper
limit found by \citetalias{herrero02}, which corresponds to one of the
lowest \mdot\ determined for this object. If we would have compared
our findings to the higher value obtained by \cite{howarth89},
\tenlac\ would be at $\Delta \mdot = -0.5$, i.e.\ the situation in
Fig.~\ref{fig:mdot_comp} would be reversed. Consequently, the large
difference for \tenlac\ shown in this figure can not be assigned to an
error in the automated method, but rather reflects our limited
understanding of this object.

All in all, we can conclude that the general agreement between mass
loss rates obtained with the automated method and ``by eye''
determinations is very good.

\section{Implications for the properties of massive stars}
\label{sec:implic}

With our automated method we have analysed a sizeable sample of early
type stars in a homogeneous way, which allows a first discussion of
the implications the newly obtained parameters may have on the mass
and modified wind-momentum luminosity relation (WLR) of massive
stars. A thorough discussion however needs to be based on a much
larger sample, therefore at this point we keep the discussion general
and the conclusions tentative.

\subsection{On the mass discrepancy}

\begin{figure}
\centering \resizebox{8.8cm}{!}{\rotatebox{270}{
  \includegraphics{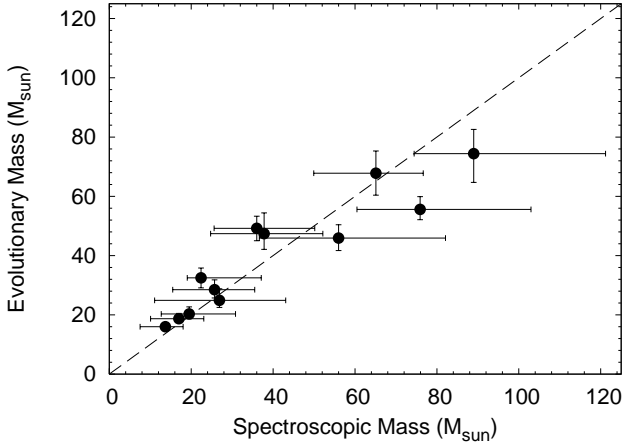}}}
  \caption{Spectroscopic masses derived in this study compared to
    evolutionary masses from \cite{schaller92}. With the gravities
    obtained from the automated fits no mass discrepancy is found and
    no systematic deviation between the spectroscopically derived
    masses and the evolutionary predicted masses can be observed.}
  \label{fig:mass_comp}
\end{figure}

The so called mass discrepancy was first noticed by
\cite{herrero92}. These authors found that the spectroscopic masses,
i.e.\ masses calculated from the spectroscopically determined gravity,
were systematically smaller than the masses predicted by evolutionary
calculations. The situation improved considerably with the use of
unified stellar atmosphere models \citep[e.g.][]{herrero02}. However,
as pointed out by \cite{repolust04} for stars with masses lower than
50~\msun\ still a milder form of a mass discrepancy appears to
persist.

Does the automated fitting method, employing the latest version of
\fastwind, help in resolving the mass discrepancy?  In
Fig.~\ref{fig:mass_comp} we present a comparison of the spectroscopic
masses calculated with the gravities obtained in this study, with
masses derived by interpolating evolutionary tracks of
\cite{schaller92}. It is clear that with the new gravities the
situation is very satisfying. All objects have spectroscopic and
evolutionary masses which agree within the error bars.

For stars with masses below 50~\msun\ a milder form of the mass
discrepancy (as found by \citetalias{repolust04}; see their Fig.~20)
could still be present, but with the present data no systematic offset
between the two mass scales can be appreciated.  Though we feel it may
be premature to conclude that the present analysis shows that the mass
discrepancy has been resolved, our results point to a clear
improvement.

\subsection{Wind-momentum luminosity relation}
\label{sec:wlr}

The modified stellar wind momentum (MWM) versus luminosity relation
offers a meaningful way to compare observed wind properties with
aspects and predictions of the theory of line driven winds (see
\citealt{kudritzki00} for a comprehensive discussion). Without going
into any detail, the modified wind momentum $\Dmom = \mdot \vinf
R_\star^{1/2}$ is predicted to be a power law of stellar luminosity.
\begin{equation}
   \log \Dmom = x \log (\lstar/\lsun) + \log D_{\circ}~,
\end{equation}
where $x$, the inverse of the slope of the line-strength distribution
function corrected for ionization effects \citep{puls00}, is expected
to be a function of spectral type and metal abundance, and $D_{\circ}$
is a function of metallicity and possibly luminosity class
\citep{markova04}. The advantageous property of \Dmom\ is that it is
not very sensitive to the stellar mass.

The limited number of stars studied in this paper is clearly
insufficient to disentangle subtleties in the \Dmom\ vs. \lstar\
relation. However, it is interesting to compare the observed and
predicted modified wind momentum, as well as to discuss the location
of \tenlac\ -- a notorious outlier.

\begin{figure}[t]
\centering
  \resizebox{8.8cm}{!}{
  \includegraphics{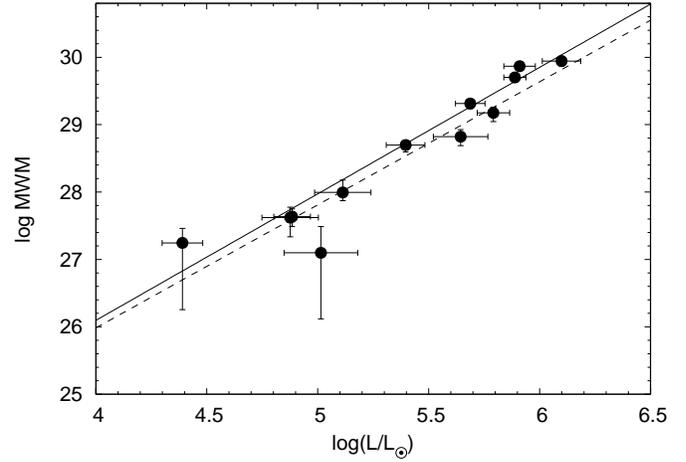}}
  \caption{Modified wind momentum (MWM) in units of
  [\mbox{g\,cm\,s$^{-2}$\rsun}] of the objects fitted with the
  automated method (solid dots). The solid line, giving the
  wind-momentum luminosity relation (WLR), corresponds to the
  regression of the modified wind momenta. Given by the dashed line is
  the predicted WLR of \cite{vink00}.}
  \label{fig:mwm_comp}
\end{figure}

Figure~\ref{fig:mwm_comp} shows this comparison between derived and
theoretical modified wind momentum. Using all programme stars to
construct an empirical linear curve in the units of this diagram gives
the following relation
\begin{equation}
  \log \Dmom = (1.88 \pm 0.09) \log (\lstar/\lsun) + (18.59 \pm 0.52)~.
\end{equation}
Within the given errors this relation is equal to the theoretical WLR
predicted by \cite{vink00}, who found $x=1.83$ and $\log \Dmom =
18.68$. Note that the low luminosity objects ($\log \lstar/\lsun
\lesssim 5.5$) also follow the average relation. Therefore, the newly
obtained mass loss rates do not show the discrepancy found by
\cite{puls96} and \cite{kudritzki00}, but confirm the work of
\citetalias{repolust04}. These authors found the low luminosity
objects to follow the general trend, based on upper limits they
obtained for the mass loss rates, whereas our new method is sensitive
enough to determine these self-consistently.

In order to investigate the effect of the anomalously low \Dmom\
obtained for \tenlac, we also constructed a WLR excluding
this object. We found that for this new relation the parameters $x$
and $\log D_{\circ}$ only changed with $\sim$0.02 and $\sim$0.01,
respectively, reflecting the large error bars found for this object.

Previous investigations by \cite{markova04} and
\citetalias{repolust04} have found the WLR to be as function of
luminosity class. Whereas the former study finds a steeper WLR for the
supergiants compared to the dwarfs, the latter finds the opposite
(though \citetalias{repolust04} remark that the subset of \cygob\
stars seem to behave more in accordance with the theoretical
result). In our sample no obvious separation is visible. In particular
note the two objects overlapping at $\log \lstar/\lsun = 4.9$ in
Fig.~\ref{fig:mwm_comp}, which are the dwarf \zoph\ and the supergiant
\cygob~\#2. To investigate a possible separation in more detail, a
separate WLR was constructed for the \cygob\ supergiants. The
resulting values of the parameters obtained are $x = 1.79 \pm 0.14$
and $\log D_{\circ} = 19.12 \pm 0.80$.  The decrease in $x$
qualitatively confirms the work of \citeauthor{markova04}. However, we
have to realize that our sample might be too small from a statistical
point of view to be able to firmly conclude whether a real separation
exists. Therefore, this question has to be postponed until we have
analysed a larger sample.

\section{Summary, conclusions and future work}
We have presented the first method for the automated fitting of
spectra of massive stars with stellar winds. In this first
implementation, a set of continuum normalized optical spectral lines
is fitted to predictions made with the fast performance non-LTE model
atmosphere code \fastwind\ by \cite{puls05}. The fitting method itself
is based on the genetic algorithm \pikaia\ by \cite{charbonneau95},
which was parallelized in order to handle the thousands of \fastwind\
models which have to be calculated for an automated fit. Concerning
the automated method we can draw the following conclusions:

\begin{enumerate}
\item [{\it i)}] The method is robust. In applying the method to a
      number of formal tests, to the study of seven O-type stars in
      \cygob, and to five Galactic stars including extreme rotators
      and/or stars with weak winds (few times $10^{-8}$ \msunyr) the
      fitting procedure did not encounter convergence problems.

\item [{\it ii)}] Using the width of the global optimum in terms of
      fitness, defining the group of best fitting models, we are able
      to define error estimates for all of the six free parameters of
      the model (\teff, \logg, helium over hydrogen abundance, \vturb,
      \mdot\ and $\beta$). These errors compare well with errors
      adopted in ``by eye'' fitting methods.

\item [{\it iii)}] For the investigated dataset our automated fitting
      method recovers mass-loss rates down to $\sim$ $6 \times 10^{-8}
      \msunyr$ to within an error of a factor of two. We point out
      that even for such low mass-loss rates it is {\em not} only the
      core of the hydrogen \ha\ line that is a mass-loss
      diagnostics. When ignoring this core the GA still recovers
      \mdot, showing that the GA is also sensitive to indirect effects
      of a change in \mdot\ on the atmospheric structure as a
      whole. However, for the method to fully take advantage of this
      information a very accurate continuum normalization is required.

\item [{\it iv)}] Though we have so far tested our method for O-type
      stars and early B-type dwarf stars, the method can also be
      applied to B and A supergiants when atomic models of diagnostic
      lines (such as \ion{Si}{iii} and \ion{Si}{iv}) are implemented
      into the analysis.

\end{enumerate}

\noindent
We have re-investigated seven O-type stars in the young cluster
\cygob\ and compared our results with the study by
\cite*{herrero02}. The \citetalias{herrero02} study uses an earlier
version of \fastwind\ and a ``by eye'' fitting procedure. The only
difference between the two studies in terms of the treatment of the
free parameters is that \citetalias{herrero02} did not treat the
microturbulent velocity and the hydrogen over helium abundance ratio
as continuous free parameters. Instead, they opted to adopt in case of
the former two possible values only. In case of the latter an initial
solar abundance was adopted which was modified in case no satisfying
fit could be obtained with this solar abundance. We have also compared
the results of an automated fitting to five early-type dwarf stars to
further investigate the robustness of our method for stars with high
rotational velocities and/or low mass loss rates. With respect to weak
winds we refer to conclusion {\it iii}. Regarding large \vsini\
values, we find that these do not pose problems for the automated
method. This is reflected in conclusion {\it i}. Concerning the
spectral analysis of the entire sample we can draw the following
conclusions:

\begin{enumerate}
\item [{\it v)}] For almost all parameters we find excellent agreement
      with the results of \citetalias{herrero02} and
      \citetalias{repolust04}, which, we note, make use of a previous
      version of \fastwind\ and an independent continuum
      normalization. The quality of our fits (in terms of fitness,
      which is a measure for the $\chi^2$ of the lines) is even better
      than obtained in these prior studies.

\item [{\it vi)}] In three cases we find a significantly higher
      surface gravity (by up to 0.36~dex). We identify two
      possible causes for this difference that may be connected to the
      difference between automated and ``by eye'' fitting. First,
      comparison of the two methods indicates that in fitting the
      Balmer line wings the latter method places essentially infinite
      strength to the observed flux at the point of maximum curvature
      of the wing profile. The automated method does not do
      this. Second, as the automated method is a multidimensional
      optimization method it may simply find a better fit to the
      overall spectrum. In at least one case this implied a higher
      temperature and significantly higher gravity.

\item [{\it vii)}] A comparison of our derived masses with those
      predicted by evolutionary calculations does not show any
      systematic discrepancy. Such a discrepancy was first noted by
      \cite{herrero92}, though was partly resolved when model
      atmospheres improved (e.g.\ see \citetalias{herrero02}). Still,
      with state-of-the-art models a mild form of a mass discrepancy
      remained for stars with masses below 50 \msun\ \citepalias[e.g.\
      see][]{repolust04}. The automated fitting approach in
      combination with the improved version of \fastwind\ does not
      find evidence for a mass discrepancy, although we remark that a
      truly robust conclusion, particularly for stars between 20 and
      50 solar masses, may require the investigation of a larger
      sample.

\item [{\it viii)}] The empirical modified wind momentum relation
      constructed on the basis of the twelve objects analysed in this
      study agree to within the error bars with the theoretical MWM
      relations based on the \cite{vink00} predictions of mass loss
      rates.
\end{enumerate}

\noindent
This first implementation of a genetic algorithm combined with the
fast performance code \fastwind\ already shows the high potential of
automatic spectral analysis. With the current rapid increase in
observations of early-type massive stars the need for an automated
fitting method is evident. We will first use our method to analyse the
$\sim$100 O-type and early B-type stars observed in the VLT large
programme {\em FLAMES Survey of Massive Stars} \citep{evans05} in the
Galaxy and the Magellanic Clouds in a homogeneous way. Future
development of the automated fitting method is likely to be in
conjunction with the further development of \fastwind. Improvements
will include the modeling of: near-infrared lines (see e.g.\
\citealt{lenorzer04} and \citealt{repolust05}), optical CNO lines (see
e.g. \citealt{trundle04}), and possibly UV resonance lines. Additional
model parameters that may be constrained within an automated approach
include a depth dependent profile for the microturbulent velocity and
small scale clumping. Within the current implementation, most likely
the method will also be able to constrain the terminal flow velocity
of A-type supergiants \citep{mccarthy97, kudritzki99}.

\acknowledgements{We would like to thank Chris Evans, Ian Hunter,
  Stephen Smartt and Wing-Fai Thi for constructive discussions, and
  Michiel Min for sharing his insights in automated fitting. M.R.M.\
  acknowledges financial support from the NWO Council for Physical
  Sciences. F.N. acknowledges PNAYA2003-02785-E and
  AYA2004-08271-C02-02 grants and the Ramon y Cajal program.}

\bibliographystyle{aa}
\bibliography{ga_ostars}

\end{document}